\documentclass[journal,twocolumn,10pt,twoside]{IEEEtran}
\ifCLASSINFOpdf
\else
\fi

\ifCLASSINFOpdf
\usepackage[pdftex]{graphicx}
\else
\usepackage[dvips]{graphicx}
\fi
\usepackage{setspace}

\usepackage{url}
\usepackage{subfigure}

\usepackage[cmex10]{amsmath}
\makeatletter
\g@addto@macro\normalsize{%
	\setlength\abovedisplayskip{2pt}
	\setlength\belowdisplayskip{2pt}
	\setlength\abovedisplayshortskip{2pt}
	\setlength\belowdisplayshortskip{2pt}
}
\makeatother
\usepackage{amssymb}
\usepackage{color}

\allowdisplaybreaks

\usepackage{array}

\usepackage{cite}
\usepackage{bbm}

\usepackage{amsmath}
\usepackage{accents}
\newlength{\dhatheight}

\newtheorem{theorem}{Theorem}
\newtheorem{definition}{Definition}

\newtheorem{proposition}{Proposition}
\newtheorem{corollary}{Corollary}

\begin{document}

%
\title{Rate-Distortion-Perception Tradeoff for \\Gaussian Vector Sources}

\author{Jingjing~Qian, Sadaf~Salehkalaibar,  Jun~Chen, Ashish Khisti, Wei Yu, \\ Wuxian~Shi,  Yiqun~Ge and Wen Tong
\thanks{Manuscript submitted to \emph{IEEE Journal on Selected Areas in Information Theory}
on December 15, 2023, revised on June 24, 2024 and November 8, 2024, accepted on November 11, 2024. This work was supported by Huawei Technologies Canada. The material in this paper has been presented in part at the IEEE International Symposium on Information Theory Workshops (ISIT-W), Athens, Greece, July 2024.}
	\thanks{Jingjing Qian and Jun Chen are with the Department of Electrical and Computer Engineering at McMaster University, Hamilton, ON L8S 4K1, Canada (email: \{qianj40,  chenjun\}@mcmaster.ca).}
 \thanks{Sadaf Salehkalaibar was with the Department of Electrical and Computer Engineering at the University of Toronto, Toronto, Canada. She is now with the Department of Computer Science at the University of Manitoba, Winnipeg, R3T 5V6, Canada (email: Sadaf.Salehkalaibar@umanitoba.ca).}
 \thanks{Ashish Khisti and Wei Yu are with the Department of Electrical and Computer Engineering at the University of Toronto, Toronto, M5S 3G4, Canada (email:\{akhisti, weiyu\}@ece.utoronto.ca).}
	\thanks{Wuxian Shi,  Yiqun Ge and Wen Tong are with   the Ottawa Research Center, Huawei Technologies, Ottawa, ON K2K 3J1, Canada (email: \{wuxian.shi, yiqun.ge, tongwen\}@huawei.com).}
	}

\maketitle

\begin{abstract}  
This paper studies the rate-distortion-perception (RDP) tradeoff for a
Gaussian vector source coding problem where the goal is to compress the
multi-component source subject to distortion and perception constraints. Specifically, the RDP setting with either the Kullback-Leibler (KL) divergence or Wasserstein-2 metric as the perception loss function is examined, and it is shown that for Gaussian vector
sources, jointly Gaussian reconstructions are optimal. We further demonstrate
that the optimal tradeoff can be expressed as an optimization problem, which
can be explicitly solved. An interesting property of the optimal solution is as
follows.  Without the perception constraint, the traditional reverse
water-filling solution for characterizing the rate-distortion (RD) tradeoff of
a Gaussian vector source states that the optimal rate allocated to each
component depends on a constant, called the water level. If the variance of a
specific component is below the water level, it is assigned a {zero}
compression rate. However, with active distortion and perception constraints,
we show that the optimal rates allocated to the different components are always
{positive}.  Moreover, the water levels that determine the optimal rate
allocation for different components are unequal. We further treat the special
case of perceptually perfect reconstruction and study its RDP function in the
high-distortion and low-distortion regimes to obtain insight to the structure
of the optimal solution. 
\end{abstract}

\section{Introduction}
The rate-distortion-perception (RDP) function is a generalization of Shannon's
rate-distortion function that incorporates an additional perception loss
function which measures the distance between the distributions of the source
and the reconstruction. It has been observed that in the neural compression
framework \cite{image-comp1, image-comp2, image-comp3, image-comp4}, improving
realism in the reconstruction comes at the price of increased distortion. In
this framework, realism is controlled by a perception loss function between the
distributions of the source and the reconstruction, while distortion is
controlled via a standard distortion loss function on the samples of the source
and its reconstruction, e.g., in terms of mean squared error.  The RDP function 
introduced in Blau and Michaeli~\cite{blau2019rethinking} formalizes this tradeoff.

The extension of classical rate-distortion (RD) theory to incorporate
constraints on the distribution of the reconstruction samples has been 
studied in various works in the information theory literature; see e.g.,~\cite{Saldi} 
and references therein. More recently, Theis and Wagner~\cite{Theis-Wagner}
present a one-shot coding theorem by means of the strong functional
representation lemma (SFRL) \cite{LiElGamal} to establish the operational validity of
the RDP function~\cite{blau2019rethinking}. In~\cite{Jun-Ashish2021}, the
authors establish analytic properties of the RDP function for the special case
of (scalar) Gaussian sources, with a quadratic distortion function and a
perception loss function of either Kullback–Leibler (KL) divergence or
Wasserstein-2 distance between the source and the reconstruction distributions.
The role of common randomness in the study of RDP function has been reported
in~\cite{wagner2022rate,Jun-JSAIT}.  Furthermore, the distortion-perception
tradeoff with a squared error distortion and Wasserstein-2 perception loss 
has been studied in~\cite{freirich2021theory,yan2021perceptual}, where it is
shown that the entire tradeoff curve can be achieved by interpolating the two
extremal reconstructions based on a given representation. Other related works
include~\cite{Huan-Liu,Jun-Ashish2023}.

This paper studies the RDP function of a Gaussian vector source under a 
squared error distortion and either KL divergence or Wasserstein-2 distance 
as the perception loss metric. Our result is thus an extension of prior
work \cite{Jun-Ashish2021} on scalar Gaussian sources to the case of vector sources. 
We start by demonstrating the optimality of jointly Gaussian reconstructions
for Gaussian vector sources in the RDP setting. We then show that by decomposing 
the Gaussian vector source using the unitary transformation obtained from the eigenvalue decomposition of its covariance matrix, 
it is possible to derive an achievable RDP function of the Gaussian vector
source in term of the RDP functions of its constituent scalar components.  
The optimality of this achievable scheme can be established by a converse
proof.  This means the characterization of the optimal RDP function 
can be formulated as an optimization problem.  We explicitly derive the
solution of the optimization problem and investigate structural properties 
of the optimal solution.

\begin{figure*}[t]
	\centering
	\includegraphics[scale=0.35]{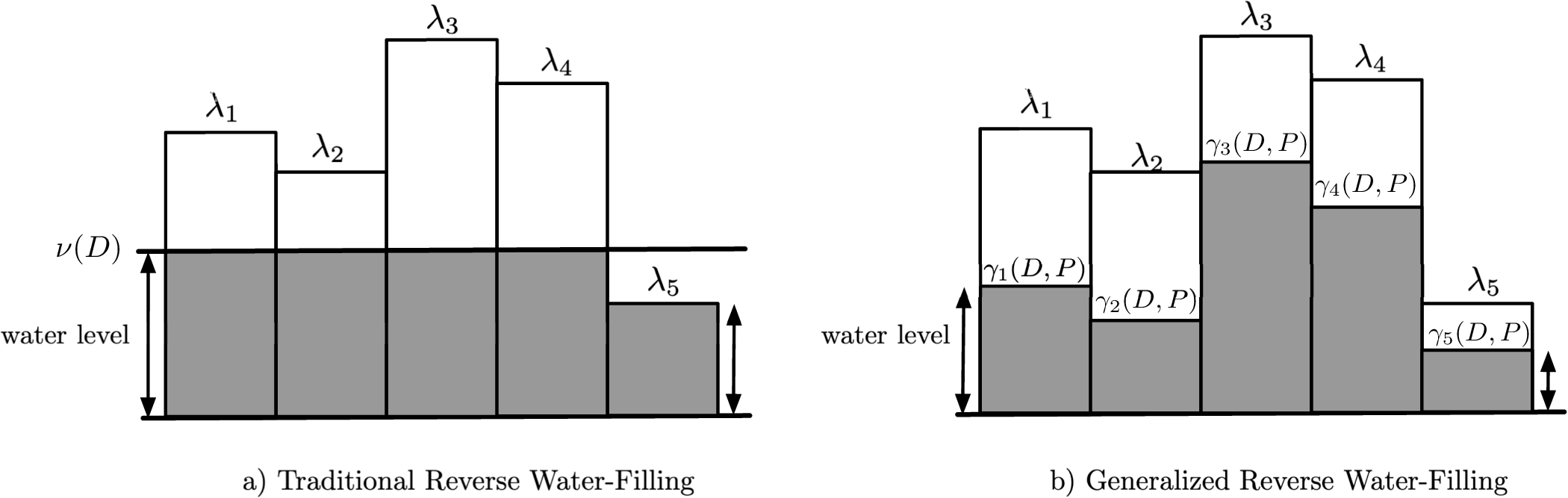}
	\caption{(a) Without a perception constraint, the traditional reverse water-filling solution for a parallel Gaussian source fixes a constant water level. When the variance of a specific component is less than the water level, it is assigned zero rate. (b) With an active perception constraint, unequal water levels are assigned to different components. The variance of each component is always greater than the corresponding water level.
	Every component has a positive rate.}
	\label{waterfilling-figure}
\end{figure*}

The optimal RDP function for the Gaussian vector source has the following 
interesting property. Without the perception constraint, the rate-distortion
function of a parallel Gaussian source model has a classical \emph{reverse
water-filling} characterization \cite[Thm 10.3]{Cover}, where the optimal rate
allocation across the components is computed according to a distortion dependent
parameter called \emph{water level}. A positive rate is assigned to those
components that have a variance above this parameter. Any component whose
variance is below the water level has a zero rate; see
Fig.~\ref{waterfilling-figure}(a). However, with a perception constraint, we
observe a qualitatively different solution as shown in
Fig.~\ref{waterfilling-figure}(b).  First, unlike the case of reverse
water-filling, the associated water level for each component can be different
and is characterized as a solution to a set of equations.  Second, while
reverse water-filling assigns zero rate to those source components whose
variances are below the water level, all components in the RDP setting are
assigned a non-zero rate as long as both distortion and perception
constraints are active.

We further consider the special case of zero perception loss (so the source and
reconstruction distributions are identical) and establish analytical results in
this case. Moreover, we present asymptotic results on high and low distortion
cases with zero perception, and shed additional insights into the difference
between the RDP function and the RD function.

The rest of the paper is organized as follows. In
Section~\ref{section:system_model}, we introduce the system model and some
preliminaries. Some basics on the traditional reverse water-filling
solution are provided in Section~\ref{section:water-filling}. We discuss
the generalized water-filling solution in Section~\ref{section:RDP} for
both KL-divergence and Wasserstein-2 distance as perception metrics; some
properties of the RDP function are also discussed for perfect perceptual
reconstruction; the asymptotic analysis is provided for both low and high
distortion regimes.

Notation: We denote entropy, differential entropy and mutual information
by $H(.)$, $h(.)$ and $I(.;.)$, respectively. The cardinality of the set
$\mathcal{X}$ is written as $|\mathcal{X}|$. We use $P_X$ to denote the 
probability distribution function of a random vector $X$. We use $\mathcal{N}(\mu,\Sigma)$
to denote the Gaussian distribution with mean $\mu$ and covariance matrix
$\Sigma$. We use $\mathbb{E}[\cdot]$ to denote the expectation operator,
and $\mathbb{R}$ to denote the set of real numbers. Throughout this paper, the base of the logarithm function is
$e$.

\section{System Model and Preliminaries}
\label{section:system_model}

Let $X\sim P_{X}$ be 
an $L$-dimensional Gaussian vector source with mean $0$ and covariance matrix $\Sigma_X \succ 0$.  Consider the eigenvalue decomposition of $\Sigma_X$ as follows:
\begin{IEEEeqnarray}{rCl}
	\Sigma_X = \Theta^T\Lambda_X\Theta,\label{eigenvalue-decomposition-definition}
\end{IEEEeqnarray}
where $\Theta$ is unitary and $\Lambda_X$ is a diagonal matrix of positive eigenvalues\footnote{ Note that if some of the eigenvalues are zero, the corresponding columns of the unitary matrix $\Theta$ can be removed, and we have a diagonal $\Lambda_X$ of lower dimension. The rest of the derivations follows the same way.} 
\begin{IEEEeqnarray}{rCl}
	\Lambda_X= \text{diag}^L(\lambda_{1},\ldots,\lambda_{L}).\label{LambdaZl}
\end{IEEEeqnarray}

We assume that there is unlimited common randomness $K\in\mathcal{K}$ shared between the encoder and the decoder. Consider the following \emph{one-shot} encoding and decoding functions where the source samples are encoded one at a time:
\begin{IEEEeqnarray}{rCl}
	f&\colon& \mathbbm{R}^L\times \mathcal{K}\to \mathcal{M},\\
	g&\colon& \mathcal{M}\times \mathcal{K}\to \mathbbm{R}^L.
\end{IEEEeqnarray}
Here, $\mathcal{M}$ denotes the set of messages. 
Let $P_{\hat{X}}$ be the distribution of the reconstruction induced by the
encoding and decoding mechanisms. In this paper, we measure distortion using 
a \emph{squared-error} loss function $d\colon \mathbbm{R}^L\times \mathbbm{R}^L\to \mathbbm{R}_{\geq 0}$ where
$d(x,\hat{x}):=\|x-\hat{x}\|^2$. 
From a perceptual perspective, for given
probability distributions $P_X$ and $P_{\hat{X}}$, we use
$\phi(P_{X},P_{\hat{X}})$ to denote the perception loss function capturing the difference
between the two distributions. For the two perception metrics that we consider in the following discussion, we have $\phi(P_{X},P_{\hat{X}})=0$ if and only if
$P_X=P_{\hat{X}}$.

The above framework is referred to as the
one-shot setting, because it compresses one sample at a time. We can also
define the setting of encoding $n$ independently and identically 
distributed (i.i.d.) samples $X^n=(X_1,\ldots,X_n)$ and reconstructing 
$\hat{X}^n=(\hat{X}_1,\ldots,\hat{X}_n)$, and consider the asymptotic setting with $n\to \infty$.

\begin{definition}[Operational RDP Functions]\label{def-operational} Let $X \sim P_X$. For given distortion-perception constraints $(D,P)$, a rate $R$ is said to be achievable if there exist encoding and decoding functions satisfying 
	\begin{IEEEeqnarray}{rCl}
		\mathbbm{E}[\ell(M)]&\leq & R,\\
		\mathbbm{E}[\|X-\hat{X}\|^2]&\leq & D,\\
		\phi(P_X,P_{\hat{X}})&\leq & P,
	\end{IEEEeqnarray}
	where $\ell(M)$ denotes the length of the message $M$ for encoding one sample. The infimum of all achievable rates $R$ is called the \emph{one-shot rate-distortion-perception (RDP) function}, denoted as $R^o(D,P)$.

	For the asymptotic setting, given distortion-perception constraints $(D,P)$, a rate $R$ is said to be achievable if there exist encoding and decoding functions such that
	\begin{IEEEeqnarray}{rCl}
		\lim_{n\to\infty}\frac{1}{n}\sum_{i=1}^n\mathbbm{E}[\|X_i-\hat{X}_i\|^2]&\leq & D,\\
		\lim_{n\to\infty}\frac{1}{n}\sum_{i=1}^n\phi(P_{X_i},P_{\hat{X}_i})&\leq & P,
	\end{IEEEeqnarray}
	with the message $M$ that encodes $X^n$ satisfying
	\begin{IEEEeqnarray}{rCl}
		\lim_{n\to\infty}\frac{1}{n}\mathbbm{E}[\ell(M)]&\leq & R.
	\end{IEEEeqnarray}
	The infimum of all achievable rates is called the \emph{asymptotic RDP function}, denoted as $R^{\infty}(D,P)$.
\end{definition}

\begin{definition}[Information RDP Function]\label{def} 
	For given $X \sim P_X$, let $\mathcal{P}_{\hat{X}|X}(D,P)$ be the set of conditional distributions $P_{\hat{X}|X}$ such that for a fixed $(D,P)$, we have 
	\begin{equation}
		\mathbbm{E}[\|X-\hat{X} \|^2]\leq D, \qquad \phi(P_{X},P_{\hat{X}})\leq P.
	\end{equation}
	The \emph{information rate-distortion-perception (RDP) function} is defined as 
	\begin{equation}
		R(D,P) = \inf_{P_{\hat{X}|X}\in \mathcal{P}_{\hat{X}|X}(D,P)} I(X;\hat{X}).\label{RDP-function-optimal}
	\end{equation}
\end{definition}


As explained in detail later, 
using the SFRL as in \cite{LiElGamal} and following similar steps to Theorem 2 and Theorem 5 in Appendix A.2 of \cite{Jun-Ashish2021}, one can show that 
	\begin{align}
		&R(D,P)\leq R^o(D,P)\leq R(D,P)+ \log(R(D,P)+1)+5, \label{one-shot-RDP}
	\end{align}
	and
	\begin{equation}
		R^{\infty}(D,P)=R(D,P).
	\end{equation}
	Consequently, 
	the one-shot operational RDP function $R^o(D,P)$ is asymptotically close to 
	the information RDP function $R(D,P)$ and the asymptotic RDP function
		$R^{\infty}(D,P)$ at high rate. 

In the rest of the paper, the perception metric $\phi(P_X,P_{\hat{X}})$ is assumed to be either the KL-divergence, i.e., 
\begin{IEEEeqnarray}{rCl}
	D(P_{\hat{X}}\|P_X)=\int_x P_{\hat{X}}(x)\log \frac{P_{\hat{X}}(x)}{P_X(x)} dx,
\label{eq:KL}
\end{IEEEeqnarray}
or the (squared) Wasserstein-2 distance, i.e.,
\begin{IEEEeqnarray}{rCl}
	W_2^2(P_X,P_{\hat{X}})=\inf \mathbbm{E}[\|X-\hat{X}\|^2],
\label{eq:wasserstein}
\end{IEEEeqnarray}
where the infimum is taken over all joint distributions of $(X,\hat{X})$ with marginals $P_X$ and $P_{\hat{X}}$.

Before characterizing the RDP function, we first review
the case of no perception constraint, which corresponds to traditional reverse water-filling for the classical rate-distortion function.

\section{Traditional Reverse Water-Filling}
\label{section:water-filling}
The classical rate-distortion theory for a parallel Gaussian source states that the optimal rate allocated to each component depends on a constant parameter, called \emph{water level}, as shown in Fig.~\ref{waterfilling-figure}(a). 
The water level also represents the distortion allowed at those components whose variances are
above the water level. 
For a given distortion $D$, let $\nu(D)$ be the solution to the equation
\begin{IEEEeqnarray}{rCl}
	\sum_{\ell=1}^L\left[\lambda_{\ell}-\nu(D)\right]^+&=&\left[\sum_{\ell=1}^L\lambda_{\ell}-D\right]^+,\label{distortion-budget}
\end{IEEEeqnarray}
where $[x]^+:=\max\{0,x\}$. Now, let 
\begin{IEEEeqnarray}{rCl}
	\gamma_{\ell}^*(D,\infty)=\left\{\begin{array}{ll}\lambda_{\ell}&\;\;\text{if}\;\;\nu(D)\geq \lambda_{\ell},\\
	\nu(D)&\;\;\text{if}\;\;\nu(D)<\lambda_{\ell}.\end{array}\right.\label{ql-def}
\end{IEEEeqnarray}
The rate-distortion function for the Gaussian vector source with variance $\lambda_\ell$ for its $\ell$-th component, $\ell\in\{1,\ldots,L\}$, is as follows.
\begin{theorem}[{\cite[Th. 10.3]{Cover}}]\label{traditional-RD} For a Gaussian vector source, we have
	\begin{IEEEeqnarray}{rCl}
		R(D,\infty) = \frac{1}{2}\sum_{\ell=1}^L\log \frac{\lambda_{\ell}}{\gamma^*_{\ell}(D,\infty)}.
		\label{eq:reverse_waterfilling}
	\end{IEEEeqnarray}
\end{theorem}
To simplify notation, we can redefine the water level as $\gamma_{\ell}^*(D,\infty)$ in order to
account for the components whose variances are below the water level. 
If $\lambda_{\ell}$ is below $\nu(D)$ for some $\ell$, then we set $\gamma_{\ell}^*(D,\infty) = \lambda_\ell$ and assign zero rate to this component. Two special cases of the above theorem are of particular interest. 
\begin{proposition}[High-Distortion Compression]\label{low-rate-large-P} In the high-distortion regime, we have that for sufficiently small $\epsilon>0$, 
	\begin{IEEEeqnarray}{rCl}
		\label{eq:RD_rate_high_distortion} 
		R\left(\sum_{\ell=1}^L\lambda_{\ell}-\epsilon,\infty\right)= \frac{\epsilon}{2\lambda^{\max} }+O(\epsilon^2),
	\end{IEEEeqnarray}
	where $\lambda^{\max}=\max_{\ell}\lambda_{\ell}$. Let $\mathcal{L}^{\max}$ denote the set of indices where their corresponding eigenvalues are equal to $\lambda^{\max}$. Then, the water levels are given by
	\begin{subequations}\label{water-level-high-D-high-P}
		\begin{IEEEeqnarray}{rCl}
			\gamma_{\ell}^*\left(\sum_{\ell=1}^L\lambda_{\ell}-\epsilon,\infty\right)&=&\lambda_{\ell},\;\; \forall \ell\in\{1,\ldots,L\}\backslash \mathcal{L}^{\max}, \\
			\gamma^*_{\ell^{\max}}\left(\sum_{\ell=1}^L\lambda_{\ell}-\epsilon,\infty\right)&=& \lambda^{\max}-\frac{\epsilon}{|\mathcal{L}^{\max}|}, \forall\ell^{\max}\in \mathcal{L}^{\max}.\nonumber\\
		\end{IEEEeqnarray}
	\end{subequations}
\end{proposition}
\begin{IEEEproof} See Appendix~\ref{low-asym}.
\end{IEEEproof}
The above proposition states that in high-distortion compression, a positive rate is only assigned to the components with the largest eigenvalue. 

\begin{proposition}[Low-Distortion Compression]\label{high-rate-large-P} In the low-distortion regime, we have that for a sufficiently small $\epsilon>0$,
	\begin{IEEEeqnarray}{rCl}
		\label{eq:RD_rate_low_distortion} 
		R(\epsilon,\infty) = \frac{1}{2}\sum_{\ell=1}^L\log \frac{L\lambda_{\ell}}{\epsilon},
	\end{IEEEeqnarray}
	where the water levels are given by
	\begin{IEEEeqnarray}{rCl}
		\gamma_{\ell}^*(\epsilon,\infty)=\frac{\epsilon}{L},\qquad \forall \ell\in\{1,\ldots,L\}.
	\end{IEEEeqnarray}
\end{proposition}
\begin{IEEEproof} See Appendix~\ref{high-asym}.
\end{IEEEproof}
For low-distortion compression, according to the above proposition, the same water level is assigned to all components.

\section{Rate-Distortion-Perception Function}
\label{section:RDP}



\subsection{Optimality of Gaussian Reconstruction}

We first present a result showing that for the two perception metrics (\ref{eq:KL}) and (\ref{eq:wasserstein}) and for a Gaussian vector source, jointly Gaussian reconstruction is optimal. 

\begin{theorem}\label{thm-Gaussian-opt} For a zero-mean Gaussian source $X$, if the perception metric is either the KL-divergence or the Wasserstein-2 distance, without loss of optimality, in the optimization problem \eqref{RDP-function-optimal},  we can restrict the reconstruction $\hat{X}$ to have mean zero and be jointly Gaussian with $X$.
\end{theorem}
\begin{IEEEproof} See Appendix~\ref{thm-Gaussian-opt-proof}.
\end{IEEEproof}

A common property of the two perception metrics that enables the above theorem
to hold is that if the source is Gaussian distributed, conditional Gaussian
reconstruction minimizes both metrics among those with the same
first- and second-order joint statistics. Theorem \ref{thm-Gaussian-opt}
implies that the optimization of RDP function can be restricted to jointly
Gaussian distributions that satisfy the distortion and perception constraints. 

\subsection{RDP Function with KL Divergence as Perception Metric}\label{section:RDP_KL}


In this section, we present the RDP function with the KL-divergence as the perception metric, i.e., $\phi(P_{X},P_{\hat{X}})=D(P_{\hat{X}}\|P_X)$. The results for the Wasserstein-2 distance as the perception metric is stated in the subsequent section. 
We present both one-shot and asymptotic RDP functions.
As already mentioned, the one-shot RDP function $R^o(D,P)$ is close to the information RDP function $R(D,P)$ at high rate. Here we provide explicit constructions of both one-shot and asymptotic coding strategies for achieving (close to)  $R(D,P)$. 


The first step is to decompose the source using eigenvalue decomposition as in~\eqref{eigenvalue-decomposition-definition} and define 
\begin{IEEEeqnarray}{rCl}
	Z=\Theta X.
\end{IEEEeqnarray}
The main idea is to construct a new Gaussian random vector $\hat{Z}$ and to use
the channel simulation result of \cite{LiElGamal} to communicate $\hat{Z}$ to 
the decoder at a rate of $R$.  The new random vector $\hat{Z}$ is
designed to be correlated with $Z$ in a very specific way in order to satisfy the distortion 
and perception constraints $D$ and $P$, respectively. 
The correlation between $Z$ and $\hat{Z}$ is controlled by two sets of parameters,
$\{\gamma_{\ell}\}_{\ell=1}^L$ and $\{\hat{\lambda}_{\ell}\}_{\ell=1}^L$, such that $0<\gamma_{\ell}\leq \lambda_{\ell}$ and $0<\hat{\lambda}_{\ell}\leq \lambda_{\ell}$. The optimal values of these parameters are determined later. 

In effect, instead of the classical rate-distortion setting where $\hat{Z}$ is
chosen to minimize the rate subject to the distortion constraint, here we choose 
$\hat{Z}$ to satisfy both distortion and perception constraints. We construct 
this noisy version of $Z$ at the decoder by taking advantage of the availability 
of common randomness. 

Specifically, $\hat{Z}$ is a zero-mean  random vector with a joint Gaussian distribution with $Z$ such that $(Z_{\ell},\hat{Z}_{\ell})$ for different $\ell\in\{1,\ldots,L\}$, are mutually independent and
\begin{align}
	&\hspace{-0.5cm}\mathrm{cov}(Z_{\ell}, \hat{Z}_{\ell}) =
	\left[
		\begin{array}{cc}
			\lambda_\ell & \sqrt{\hat{\lambda}_{\ell}(\lambda_{\ell}-\gamma_{\ell})} \\
			\sqrt{\hat{\lambda}_{\ell}(\lambda_{\ell}-\gamma_{\ell})} & \hat{\lambda}_\ell 
		\end{array}
		\right].\label{eq:covariance_Z_hat}
\end{align}
With the above covariance structure, we can verify that 
$\gamma_{\ell}$ is the minimum mean-squared error (MMSE) of estimating $Z_\ell$ based on $\hat{Z}_\ell$, i.e.,
\begin{IEEEeqnarray}{rCl}
	\gamma_{\ell}=\mathbbm{E}[(Z_{\ell}-\mathbbm{E}[Z_{\ell}|\hat{Z}_{\ell}])^2].
\end{IEEEeqnarray}

Now, to derive an upper bound on the one-shot RDP function $R^o(D,P)$, we can use
a consequence of the SFRL \cite[Theorem 1]{LiElGamal} 
to show that when common randomness $K$ is available at both the encoder and decoder, there exists a channel simulation scheme that allows $\hat{Z}_\ell$ to be reconstructed at the decoder at a communication rate of
\begin{IEEEeqnarray}{rCl}
	I(Z_{\ell};\hat{Z}_{\ell})+\log(I(Z_{\ell};\hat{Z}_{\ell})+1)+5.
\end{IEEEeqnarray}
After the reconstruction of $\hat{Z}_{\ell}$ at the decoder, we use the same unitary matrix to transform it into $\hat{X}$, i.e., 
\begin{IEEEeqnarray}{rCl}
	\hat{X}=\Theta^T \hat{Z}.
\end{IEEEeqnarray}
The above scheme leads to the one-shot rate, distortion, and perception loss for the $\ell$-th component of $Z$ as functions of $\lambda_\ell$, $\hat{\lambda}_\ell$ and $\gamma_\ell$ as follows:
\begin{IEEEeqnarray}{rCl}
	\hspace{-1cm}{R}_{\ell}( \gamma_{\ell})&=& \frac{1}{2}\log\left(\frac{\lambda_{\ell}}{\gamma_{\ell}}\right)+\log \left(\frac{1}{2}\log\left(\frac{\lambda_{\ell}}{\gamma_{\ell}}\right)+1\right)  +5,\label{Rl-function}\nonumber\\\\	
	{D}_{\ell}(\gamma_\ell,\hat{\lambda}_{\ell})&=&\lambda_{\ell}-2\sqrt{\hat{\lambda}_{\ell}(\lambda_{\ell}-\gamma_{\ell})}+\hat{\lambda}_{\ell},\label{Dl-function}\\
	{P}_{\ell}(\hat{\lambda}_{\ell})  &=& \frac{1}{2}\left( \frac{\hat{\lambda}_{\ell}}{\lambda_{\ell}}-1+\log \frac{\lambda_{\ell}}{\hat{\lambda}_{\ell}}\right).\label{Pl-function}
\end{IEEEeqnarray}
This allows a characterization of an achievable one-shot RDP function of a Gaussian vector source as an optimization problem over $\hat{\lambda}_\ell$ and $\gamma_\ell$ across its components.


For the asymptotic setting, the achievable scheme is identical,
except that we compress a block of $n$ samples together. As $n \rightarrow \infty$, 
the logarithm and the constant terms in ~\eqref{Rl-function} can be neglected. This leads to an upper bound for $R^\infty(D,P)$, which is equal to $R(D,P)$. This upper bound turns out to be tight, i.e., a converse can be proved. This gives the following characterization of $R(D,P)$.

\begin{theorem}\label{KL-thm} 
	The rate-distortion-perception function $R(D,P)$ for a Gaussian vector source with parameters defined by (\ref{eigenvalue-decomposition-definition}) and (\ref{LambdaZl}), and with KL-divergence as the perception metric, is given by the solution to the following optimization problem:
	\begin{subequations}\label{thm-program}
		\begin{IEEEeqnarray}{rCl}
			R(D,P)= & \min_{\{\hat{\lambda}_{\ell},\gamma_{{\ell}}\}_{\ell=1}^L} & \frac{1}{2}\sum_{\ell=1}^L \log \frac{\lambda_{\ell}}{\gamma_{{\ell}}}\label{obj-func-program-thm}\\
			&\text{s.t.} & 0<\gamma_{\ell}\leq \lambda_{\ell} \label{constraint1-program-thm}\\
			&&0\leq\hat{\lambda}_{\ell} \label{constraint2-program-thm}\\
			&&\sum_{\ell=1}^L {D}_{\ell}(\gamma_{\ell},\hat{\lambda}_{\ell}) \leq D,\label{constraint3-program-thm}\\
			&&\sum_{\ell=1}^L {P}_{\ell}( \hat{\lambda}_{\ell})\leq P.\label{constraint4-program-thm}
		\end{IEEEeqnarray}
	\end{subequations}
\end{theorem}
\begin{IEEEproof} See Appendix~\ref{KL-app}.
\end{IEEEproof}

An interpretation of the above is as follows.
For a given $(D,P)$, let $\gamma_{\ell}^*(D,P)$ and $\hat{\lambda}^*_{\ell}(D,P)$, 
$\ell\in\{1,\ldots,L\}$, be the optimal solution to (\ref{thm-program}). 
Comparing this with (\ref{eq:reverse_waterfilling}), it can be seen that 
$\gamma^*_{\ell}(D,P)$ can be interpreted as the water level for the $\ell$-th 
component, which determines the rate allocated to that component according
to \eqref{obj-func-program-thm}; see Fig.~\ref{waterfilling-figure}(b). 

\subsection{Generalized Water-filling with KL Divergence as Perception Metric}

We now proceed to analyze the solution to the optimization program in Theorem~\ref{KL-thm}.
It can be shown that the optimization problem (\ref{thm-program})
is convex. 
Let $\nu_1$, $\nu_2$, $\{\xi_{\ell}\}_{\ell=1}^L$, $\{\eta_{\ell}\}_{\ell=1}^L$ be nonnegative Lagrange multipliers. For $\ell\in\{1,\ldots,L\}$, we have the first-order conditions:
\begin{equation}
	\frac{1}{2\gamma^*_{{\ell}}(D,P)}-\nu_1\sqrt{\frac{\hat{\lambda}^*_{{\ell}}(D,P)}{\lambda_{{\ell}}-\gamma^*_{{\ell}}(D,P)}}-\xi_{\ell}=0,\label{cond1-opt}
\end{equation}
and
\begin{align}
	&\nu_1\left(-\sqrt{\frac{\lambda_{{\ell}}-\gamma^*_{{\ell}}(D,P)}{\hat{\lambda}^*_{{\ell}}(D,P)}}+1\right)+\nonumber\\&\hspace{2cm}\frac{\nu_2}{2}\left(\frac{1}{\lambda_{{\ell}}}-\frac{1}{\hat{\lambda}^*_{{\ell}}(D,P)}\right)
	-\eta_{\ell}=0.\label{cond2-opt}
\end{align}
We first focus on the most interesting regime where the distortion and the perception constraints are both active so $\nu_1,\nu_2>0$, and $\gamma_\ell < \lambda_\ell$, $\hat{\lambda}_\ell > 0$ so that
$\xi_{\ell}=\eta_{\ell}=0$ for all $\ell\in\{1,\ldots,L\}$. 
In this case, \eqref{cond1-opt} implies
that $\hat{\lambda}^*_{\ell}(D,P)$ can be expressed as 
\begin{IEEEeqnarray}{rCl}
	\hat{\lambda}_{\ell}^*(D,P)=\frac{\lambda_{\ell}-\gamma^*_{\ell}(D,P)}{4\gamma^{*2}_{\ell}(D,P)\nu_1^2}.\label{eq:lambda_hat_star_KL}
	\end{IEEEeqnarray}
Together with \eqref{cond2-opt}, this means that 
$\gamma_{\ell}^*(D,P)$ is the positive solution to the following equation
\begin{IEEEeqnarray}{rCl}
	&&\nu_1(1-2\nu_1\gamma^*_{\ell}(D,P))=\frac{1}{2}\nu_2\left(\frac{4\gamma^{*2}_{\ell}(D,P)\nu_1^2}{\lambda_{\ell}-\gamma^*_{\ell}(D,P)}-\frac{1}{\lambda_{\ell}}\right),\;\;\;\;
	\end{IEEEeqnarray}
which is quadratic in $\gamma^*_{\ell}(D,P)$ and can be solved analytically as follows: 
\begin{IEEEeqnarray}{rCl}
	\label{eq:gamma_star_KL}
	&&\gamma^*_{\ell}(D,P)=\frac{-2\lambda_{\ell}\nu_1(1+2\lambda_{\ell}\nu_1)-\nu_2+\sqrt{\Delta(\nu_1,\nu_2)}}{8\lambda_{\ell}\nu_1^2(-1+\nu_2)},\nonumber\\
\end{IEEEeqnarray}
where 
\begin{align}
\Delta(\nu_1,\nu_2)&= (\nu_2+2\lambda_{\ell}\nu_1+4\lambda_{\ell}^2\nu_1^2)^2\nonumber\\&\hspace{0.5cm}+16\lambda_{\ell}^2\nu_1^2(\nu_2+2\lambda_{\ell}\nu_1)(\nu_2-1).
\end{align}
There is an alternative expression for $\gamma^*_{\ell}(D,P)$ in term of $\hat{\lambda}_{\ell}^*(D,P)$ that can be obtained by solving (\ref{eq:lambda_hat_star_KL}) as a quadratic equation in $\gamma^*_{\ell}(D,P)$ as below:
\begin{equation}
	\gamma^*_{\ell}(D,P)=\frac{2\lambda_{\ell}}{1+\sqrt{1+16\lambda_{\ell}\hat{\lambda}_{\ell}^{*}(D,P)\nu_1^2}}.\label{gamma-lambdahat}
\end{equation}
This expression is useful later in Corollary \ref{0-PLF-cor}.

The expressions \eqref{eq:gamma_star_KL} and \eqref{eq:lambda_hat_star_KL}
give us the following generalized reverse water-filling
interpretation of the optimal RDP solution. At given distortion constraint $D$
and perception constraint $P$, each component of the source with variance
$\lambda_\ell$ is reconstructed by $\hat{Z}_\ell$ having a variance
$\hat\lambda^*_\ell(D,P)$. Because $\gamma^*_\ell(D,P)$ is the variance of the MMSE 
estimate of $Z_\ell$ given $\hat{Z}_\ell$, this requires a rate of
$\frac{1}{2}\log\left(\frac{\lambda_\ell}{\gamma^*_\ell(D,P)}\right)$.  
The parameters $\hat\lambda^*_\ell(D,P)$ and $\gamma^*_\ell(D,P)$ are chosen to
satisfy the distortion and perception constraints. As already mentioned, $\gamma^*_\ell(D,P)$ can 
be thought of as the water level, cf. (\ref{eq:reverse_waterfilling}).

When both the distortion and the perception constraints are active,
i.e., $\nu_1, \nu_2 > 0$, 
it is possible to prove (as shown in the theorem below) that 
\begin{equation}
	\label{eq:positive_rate}
	\gamma^*_\ell(D,P) < \lambda_\ell, \quad \forall \ell \in \{1,\cdots,L\}, 
\end{equation}
so every component of the source is always allocated a non-zero rate regardless of the distortion 
constraint---unlike the traditional reverse water-filling solution, where a component may be 
allocated zero rate if its variance is below the water level. Moreover,  in contrast to the traditional reverse water-filling solution, the distortion of each component (i.e., $D_{\ell}(\gamma_{\ell}^*(D,P),\hat{\lambda}_{\ell}^*(D,P))$) may not be the same across the different components. So, the optimal distortion allocation across the components may be unequal when both perception and distortion constraints are active. 
 
It is also possible that either the distortion or the perception constraint is not active. If the distortion constraint is active while the perception constraint is inactive, i.e., $\nu_1>0$ and $\nu_2=0$, and $\eta_{\ell}=\eta'_{\ell}=0$ for all $\ell\in\{1,\ldots,L\}$, then~\eqref{cond1-opt} and~\eqref{cond2-opt} yield the traditional reverse water-filling solution. Specifically, the water level is given by $\min\{\frac{1}{2\nu_1},\lambda_{\ell}\}$ where $\nu_1$ satisfies the following:
\begin{IEEEeqnarray}{rCl}
\sum_{\ell=1}^L\left[\lambda_{\ell}-\frac{1}{2\nu_1}\right]^+=\left[\sum_{\ell=1}^L\lambda_{\ell}-D\right]^+.
\end{IEEEeqnarray}
By redefining $\frac{1}{2\nu_1}$ as $\nu(D)$, we see that the above expression is the same as \eqref{distortion-budget}. 

If the distortion constraint is inactive, i.e., $\nu_1=0$, based on~\eqref{cond1-opt}, we have $\xi_{\ell}>0$ which yields 
\begin{equation}
	\label{eq:zero_rate}
\gamma^*_{\ell}(D,P)=\lambda_{\ell},\qquad \forall\ell\in\{1,\ldots,L\}.
\end{equation} 
This implies that every component of the source is assigned a zero rate if the distortion constraint is not active. The decoder simply generates the reconstruction independent of the source using a distribution that satisfies the perception constraint. Such a distribution may not be unique, as shown in the theorem below.

An interesting observation is that based on 
(\ref{eq:positive_rate}) and (\ref{eq:zero_rate}), we see that 
when the perception constraint is active, 
it is either that all the components are allocated positive rate, 
or that all the components are allocated zero rate. 
This means that the situation in the traditional reverse water-filling, where some of the water levels are below the eigenvalues while others are equal to the eigenvalues, cannot happen, when the perception constraint is active. 

The above discussion is summarized in the following.

\begin{theorem}\label{thm-optimal-sol} Let $(D,P)$ be strictly feasible distortion and perception constraints. The optimal solution of~\eqref{thm-program} with KL divergence as the perception metric is given as follows:
\begin{enumerate}
	\item If both the distortion and perception constraints are active\footnote{A constraint of a minimization problem is said to be inactive if the optimization problem with the same objective function but with the said constraint removed (while keeping all the other constraints) has at least one optimal solution that already satisfies all the original constraints.}, then there exist
	$\nu_1, \nu_2 > 0$ such that
	$\gamma^*_{\ell}(D,P)$ is as expressed in (\ref{eq:gamma_star_KL}) and 
	$\hat{\lambda}^*_{\ell}(D,P)$ is as expressed in (\ref{eq:lambda_hat_star_KL}). Here, $\nu_1$ and $\nu_2$ are chosen such that
	\begin{IEEEeqnarray}{rCl}
		\hspace{-0.5cm}\sum_{\ell=1}^L D_{\ell}(\gamma^*_{\ell}(D,P),\hat{\lambda}_{\ell}^*(D,P))&=& D,\label{thm-first-clause-D-constraint}\\
		\hspace{-0.5cm}\sum_{\ell=1}^L P_{\ell}(\hat{\lambda}_{\ell}^*(D,P))&=&P.\label{thm-first-clause-P-constraint}
		\end{IEEEeqnarray}
	In this case, every component has a positive rate.
\item If the distortion constraint is active but the perception constraint is inactive, then
	there exists $\nu_1 >0$ such that
	$\gamma^*_{\ell}(D,P) = \min\{\frac{1}{2\nu_1},\lambda_{\ell}\}$,
	$\hat{\lambda}^*_{\ell}(D,P) = \lambda_{\ell}-\min\{\frac{1}{2\nu_1},\lambda_{\ell}\}$ and
	\begin{IEEEeqnarray}{rCl}
		\hspace{-0.7cm}\sum_{\ell=1}^L\left[\lambda_{\ell}-\frac{1}{2\nu_1}\right]^+=\left[\sum_{\ell=1}^L\lambda_{\ell}-D\right]^+.
	\end{IEEEeqnarray}
	In this case, some components may have zero rate.
\item If the distortion constraint is inactive, then $\gamma^*_{\ell}(D,P) = \lambda_\ell$,
	and $\hat{\lambda}^*_{\ell}(D,P)$ can be any value in the set 
	\begin{align}
		&\Bigg\{\hat\lambda_\ell  \Bigg|  \sum_{\ell=1}^L{P}_{\ell}(\hat{\lambda}_\ell) \leq  P, 
		\sum_{\ell=1}^L\lambda_{\ell}+\hat{\lambda}_{\ell} \leq D,
		\hat\lambda_\ell \ge 0  \Bigg\}.
		\end{align}
	In this case, every component has zero rate.
\end{enumerate}

\end{theorem}

\begin{IEEEproof} See Appendix~\ref{thm-optimal-sol-proof}.
\end{IEEEproof}


\subsection{RDP Function and Generalized Reverse Water-filling with Wasserstein-2 Distance as Perception Metric}

Next, consider the Wasserstein-2 distance as the perception metric, i.e.,  $\phi(P_X,P_{\hat{X}})=W_2^2(P_X,$ $P_{\hat{X}})$. To that end, we have the following definitions for distortion and perception loss functions. Let the distortion loss function of the $\ell$-th component be as in~\eqref{Dl-function}. Replace the perception loss function in \eqref{Pl-function} by the following:
\begin{IEEEeqnarray}{rCl}
{P}_{\ell}(\hat{\lambda}_{\ell})=\left(\sqrt{\lambda_{\ell}}-\sqrt{\hat{\lambda}_{\ell}}\right)^2.\label{Pl-W2-function}
\end{IEEEeqnarray}
The following theorem characterizes the RDP function with Wasserstein-2 perception loss in terms of an optimization problem.
\begin{theorem}\label{W2-thm} The rate-distortion-perception function $R(D,P)$ with Wasserstein-2 distance as the perception metric is given by the optimization program in~\eqref{thm-program} with the perception loss function (\ref{Pl-function}) replaced by \eqref{Pl-W2-function}. 
\end{theorem}
\begin{IEEEproof} The proof is similar to that of Theorem~\ref{KL-thm} with some differences which are highlighted in Appendix~\ref{W2-app}.
\end{IEEEproof}

Similar to the KL-divergence case, the optimization program for the Wasserstein-2 distance is convex. For $\ell\in\{1,\ldots,L\}$, we have the following first-order conditions:
\begin{IEEEeqnarray}{rCl}
&&\frac{1}{2\gamma_{\ell}^*(D,P)}-\nu_1\sqrt{\frac{\hat{\lambda}^*_{\ell}(D,P)}{\lambda_{\ell}-\gamma^*_{\ell}(D,P)}}-\xi_{\ell}=0,\;\;\;\;\;\;\label{cond1-opt-W2}
\end{IEEEeqnarray}
and 
\begin{IEEEeqnarray}{rCl}
&&\nu_1\left(-\sqrt{\frac{\lambda_{{\ell}}-\gamma^*_{{\ell}}(D,P)}{\hat{\lambda}^*_{{\ell}}(D,P)}}+1\right)\nonumber\\&&\hspace{1cm}+\nu_2\left(1-\sqrt{\frac{\lambda_{\ell}}{\hat{\lambda}^*_{\ell}(D,P)}}\right)\label{cond2-opt-W2}
	+\eta_{\ell}=0.
\end{IEEEeqnarray}
Consider the case where both distortion and perception constraints are active, i.e., $\nu_1,\nu_2>0$ and $\xi_{\ell}=\eta_{\ell}=0$ for all $\ell\in \{1,\ldots,L\}$. In this case,~\eqref{cond1-opt-W2} and~\eqref{cond2-opt-W2} yield the following solutions
\begin{IEEEeqnarray}{rCl}
	\gamma_{\ell}^*(D,P)&=& \frac{\theta_{\ell}}{2\nu_1},\label{optimal-solution-gamma-W2}\\
	\hat{\lambda}_{\ell}^*(D,P)&=& \frac{\lambda_{\ell}}{\left(1+\frac{(1-\theta_{\ell})\nu_1}{\nu_2}\right)^2},\label{optimal-solution-lambda-W2}
	\end{IEEEeqnarray}
where $\theta_{\ell}$ is defined to be the unique solution of the following equation:
\begin{IEEEeqnarray}{rCl}
\frac{\theta_{\ell}}{1+\frac{(1-\theta_{\ell})\nu_1}{\nu_2}}=\sqrt{1-\frac{\theta_{\ell}}{2\nu_1\lambda_{\ell}}}.
\end{IEEEeqnarray}

As in the case of KL divergence, it is possible to prove that
when both the distortion and the perception constraints are active 
we have $\gamma^*_{\ell}(D,P) < \lambda_{\ell}$.
Thus, every component is compressed at a positive rate. 

When the distortion constraint is active but the perception constraint is not active, 
the problem reduces to traditional reverse water-filling.
Finally, when the distortion constraint is not active, i.e., $\nu_1=0$, 
a zero rate is assigned to all components. 
This discussion is summarized in the following.

\begin{theorem}\label{cor-W2-optimal-sol}  Let $(D,P)$ be given distortion and perception constraints that are strictly feasible.  The optimal solution of~\eqref{thm-program} with the perception metric \eqref{Pl-function} replaced by \eqref{Pl-W2-function} is given as follows:
\begin{enumerate}
	\item If both the distortion and perception constraints are active, then there exist
	$\nu_1, \nu_2 > 0$ such that
	$\gamma^*_{\ell}(D,P)$ is as expressed in \eqref{optimal-solution-gamma-W2} and 
	$\hat{\lambda}^*_{\ell}(D,P)$ is as expressed in \eqref{optimal-solution-lambda-W2}. Here, $\nu_1$ and $\nu_2$ are chosen such that
	\begin{IEEEeqnarray}{rCl}
		\sum_{\ell=1}^L D_{\ell}(\gamma^*_{\ell}(D,P),\hat{\lambda}_{\ell}^*(D,P))&=& D,\;\;\;\label{thm-first-clause-D-constraint-W2}\\
		\sum_{\ell=1}^L P_{\ell}(\hat{\lambda}_{\ell}^*(D,P))&=&P.\;\;\;\label{W2-perception-constraint-total}
	\end{IEEEeqnarray}
	In this case, every component has a positive rate. 
	\item If the distortion constraint is active but the perception constraint is inactive, then
	there exists $\nu_1 >0$ such that
	$\gamma^*_{\ell}(D,P) = \min\{\frac{1}{2\nu_1},\lambda_{\ell}\}$,
	$\hat{\lambda}^*_{\ell}(D,P) = \lambda_{\ell}-\min\{\frac{1}{2\nu_1},\lambda_{\ell}\}$ and
	\begin{IEEEeqnarray}{rCl}
		\sum_{\ell=1}^L\left[\lambda_{\ell}-\frac{1}{2\nu_1}\right]^+=\left[\sum_{\ell=1}^L\lambda_{\ell}-D\right]^+.\;\;\;\;\;\;\;
	\end{IEEEeqnarray}
	In this case, some components may have zero rate.
	\item If the distortion constraint is inactive, then $\gamma^*_{\ell}(D,P) = \lambda_\ell$, and $\hat{\lambda}^*_{\ell}(D,P)$ can be any value in the set 
	\begin{align}
		&\Bigg\{\hat\lambda_\ell  \Bigg|  \sum_{\ell=1}^L{P}_{\ell}(\hat{\lambda}_\ell) \leq  P, 
		\sum_{\ell=1}^L\lambda_{\ell}+\hat{\lambda}_{\ell} \leq D,
		\hat\lambda_\ell \ge 0  \Bigg\}.
	\end{align}
	In this case, every component has zero rate.	
\end{enumerate}	
\end{theorem}
\begin{IEEEproof} See Appendix~\ref{cor-W2-optimal-sol-app}.
\end{IEEEproof}

\subsection{Perceptually Perfect Reconstruction}

In this section, we focus on the special case of perfect perceptual quality, 
and study the properties of the RDP function with $P=0$.

\begin{figure*}[t]
	\centering
	\includegraphics[scale=0.3]{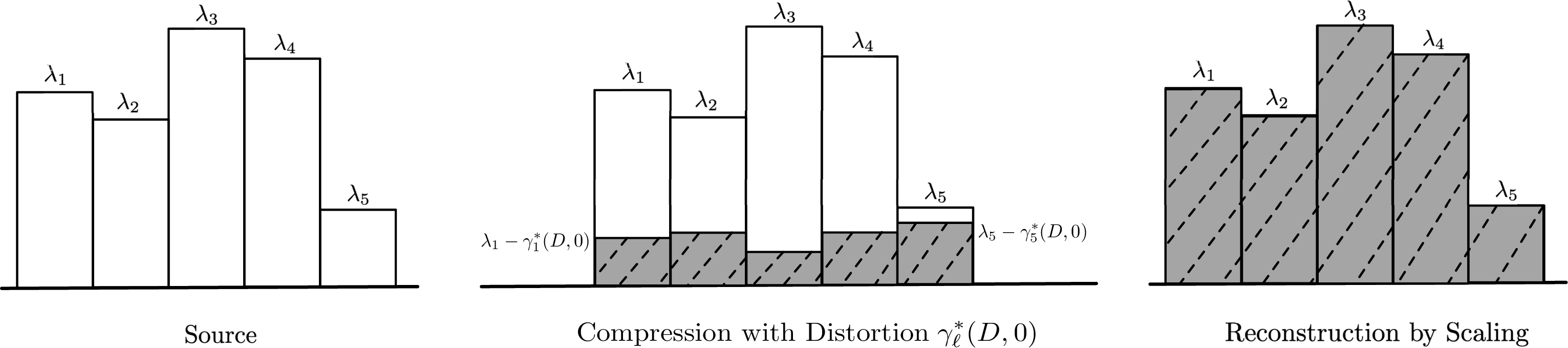}
	\caption{Generalized reverse water-filling solution for the perceptually perfect reconstruction. The source is first compressed to a representation whose components have distortion levels $\gamma_{\ell}^*(D,0)$, $\ell=1,\cdots,L$. After compression, each component has a variance given by $\lambda_{\ell}-\gamma^*_{\ell}(D,0)$. Each component is then scaled to generate a reconstruction whose distribution matches that of the original source.}
	\label{fig:P0_reconstruction}
\end{figure*}

\begin{corollary}\label{0-PLF-cor} The RDP function of a Gaussian vector source with $P=0$ is 
	\begin{IEEEeqnarray}{rCl}
		R(D,0)=\frac{1}{2}\sum_{\ell=1}^L\log \frac{1+\sqrt{1+16\nu_1^2\lambda_{\ell}^2}}{2},\;\;\label{R-def}
	\end{IEEEeqnarray}
	for some positive $\nu_1$ that satisfies 
	\begin{IEEEeqnarray}{rCl}
		D= \sum_{\ell=1}^L\left[2\lambda_{\ell}-
			2\sqrt{\lambda_{\ell}\left(\lambda_{\ell}-\gamma^*_{\ell}(D,0)\right)}\right], \;\;\;\;\label{eq:distortion-0}
	\end{IEEEeqnarray}
	where 
	\begin{IEEEeqnarray}{rCl}
		&&\gamma^*_{\ell}(D,0)=\frac{2\lambda_{\ell}}{1+\sqrt{1+16\nu_1^2\lambda_{\ell}^2}}, \;\;\;\ell\in\{1,\ldots,L\}.\label{water-level-0}
	\end{IEEEeqnarray}
\end{corollary}

\begin{IEEEproof} See Appendix~\ref{extreme-points}.
\end{IEEEproof}


An interpretation of the optimal rate allocation in this $P=0$ case is as follows.
By \eqref{R-def}, the optimal rate allocated to the $\ell$-th component is controlled by the expression $\frac{1+\sqrt{1+16\nu_1^2\lambda_{\ell}^2}}{2}$. So, if a component has a larger variance, it is compressed at a higher rate. 
Further, by \eqref{water-level-0} it also has a higher water level. 

Under general perception and distortion constraints, the encoding and decoding 
strategy adopted in this paper (which involves constructing $\hat{Z}_\ell$ as in 
\eqref{eq:covariance_Z_hat}) can be thought of as first compressing
each component of the source at an individual rate specified by the
distortion level $\gamma_\ell^*(D,P)$ based on the conventional
rate-distortion tradeoff, then scaling the compressed source to a
variance of $\hat{\lambda}_\ell^*(D,P)$ to satisfy the perception constraint.
For the perfect perception case with $P=0$, the compression rate becomes
(\ref{R-def}) and the distortion level becomes (\ref{water-level-0});  
further, each component of the compressed signal is simply scaled to match 
the variance of the source in order to ensure zero perception loss. 
The distortion after scaling is given by (\ref{eq:distortion-0}).
This is shown in Fig.~\ref{fig:P0_reconstruction}. 

We further note that at a fixed $R$, the rate allocated to each component is in general different for different $(D,P)$ tradeoff points. 
Whereas for the scalar Gaussian source, 
a \emph{universal representation} for different $(D,P)$ points at a fixed $R$ is possible via scaling \cite{Jun-Ashish2021}, for the Gaussian vector source such universal representation does not exist, due to the different rate allocations in each component at different $(D,P)$ tradeoff points.




Next, we investigate the asymptotic behavior of the compression rate and the distortion level in the perfect perception case.

\begin{proposition}[High-Distortion Compression]\label{low-rate-0-PLF} In the high-distortion and perfect perception regime, we have that for sufficiently small $\epsilon>0$,
	\begin{IEEEeqnarray}{rCl}
		R\left(2\sum_{\ell=1}^L\lambda_{\ell}-\epsilon,0\right)=\frac{\epsilon^2}{8\sum_{\ell=1}^L\lambda_{\ell}^2}+O(\epsilon^3),\;\;\;\;\;\;
	\end{IEEEeqnarray}
	where the water levels are given by
	\begin{IEEEeqnarray}{rCl}
		\gamma^*_{\ell}\left(2\sum_{\ell=1}^L\lambda_{\ell}-\epsilon,0\right)&=&\lambda_{\ell}-\frac{\epsilon^2\lambda_{\ell}^3}{4\left(\sum_{\ell=1}^L\lambda_{\ell}^2\right)^2}+O(\epsilon^3),\nonumber\\&&\hspace{2cm} \ell\in\{1,\ldots,L\}.\label{water-level-high-D-0-PLF}
	\end{IEEEeqnarray}
\end{proposition}
\begin{IEEEproof} See Appendix~\ref{low-0-PLF-app}.\end{IEEEproof}



Here, we express $R(D,0)$ in term of deviation from the maximum distortion at perfect perception at zero rate. 
This maximum distortion can be shown to be $2\sum_{\ell=1}^L\lambda_{\ell}$, which is twice of the total variance of the source \cite{Jun-Ashish2021},
because at zero rate the decoder should simply generate an independent Gaussian random vector with the same covariance matrix.
Comparing $R\left(2\sum_{\ell=1}^L\lambda_{\ell}-\epsilon,0\right)$ in
Proposition~\ref{low-rate-0-PLF} with
$R\left(\sum_{\ell=1}^L\lambda_{\ell}-\epsilon,\infty\right)$ in
Proposition~\ref{low-rate-large-P}, it is interesting to see that 
the variances of the source appear in 
$R\left(2\sum_{\ell=1}^L\lambda_{\ell}-\epsilon,0\right)$ as
$\sum_{\ell=1}^L\lambda^2_{\ell}$ which is the sum of the square of the variances over
all the components. This is in contrast to the corresponding
factor in $R\left(\sum_{\ell=1}^L\lambda_{\ell}-\epsilon,\infty\right)$ in the
traditional reverse water-filling solution which is simply
$\lambda^{\max}$. This is a consequence of the perfect perception constraint, 
which requires all the components to be reconstructed with the same variances as
the source at the decoder.


\begin{proposition}[Low-Distortion Compression]\label{high-rate-0-PLF} In the low-distortion and perfect perception regime, we have that for sufficiently small $\epsilon>0$,
	\begin{IEEEeqnarray}{rCl}
		R(\epsilon,0)=\frac{1}{2}\sum_{\ell=1}^L\log \frac{L\lambda_{\ell}}{\epsilon}+
		\frac{\epsilon}{8L}\sum_{\ell=1}^L\frac{1}{\lambda_{\ell}}+O(\epsilon^2),\nonumber\\
	\end{IEEEeqnarray}
	where the water levels are given by 
	\begin{IEEEeqnarray}{rCl}
		\gamma^*_{\ell}(\epsilon,0)&=&\frac{\epsilon}{L}-\frac{\epsilon^2}{2L^2\lambda_{\ell}}+\frac{\epsilon^2}{4L^3}\sum_{\ell=1}^L\frac{1}{\lambda_{\ell}}+O(\epsilon^3),\nonumber\\&&\hspace{3.5cm} \ell\in\{1,\ldots,L\}.\label{low-D-P0-water-level}
	\end{IEEEeqnarray}
\end{proposition}
\begin{IEEEproof} See Appendix~\ref{high-0-PLF-app}.
\end{IEEEproof}

Comparing Proposition~\ref{high-rate-0-PLF} with Proposition~\ref{high-rate-large-P}, we see that in this high-rate low-distortion regime, the extra rate required to satisfy zero-perception scales as 
\begin{IEEEeqnarray}{rCl}
	&&R(\epsilon,0)-R(\epsilon,\infty)=
	\frac{\epsilon}{8L}\sum_{\ell=1}^L\frac{1}{\lambda_{\ell}}+O(\epsilon^2),\;\;\;\;\\
	&&\gamma^*_{\ell}(\epsilon,\infty)-\gamma^*_{\ell}(\epsilon,0)=\frac{\epsilon^2}{2L^2\lambda_{\ell}}-\frac{\epsilon^2}{4L^3}\sum_{\ell=1}^L\frac{1}{\lambda_{\ell}}+O(\epsilon^3),\nonumber\\&&\hspace{5.3cm} \ell\in\{1,\ldots,L\}.
\end{IEEEeqnarray}


\begin{figure*}[t]
\centering
\subfigure[High distortion; no perception constraint] 
{
  \includegraphics[width=0.3\textwidth]{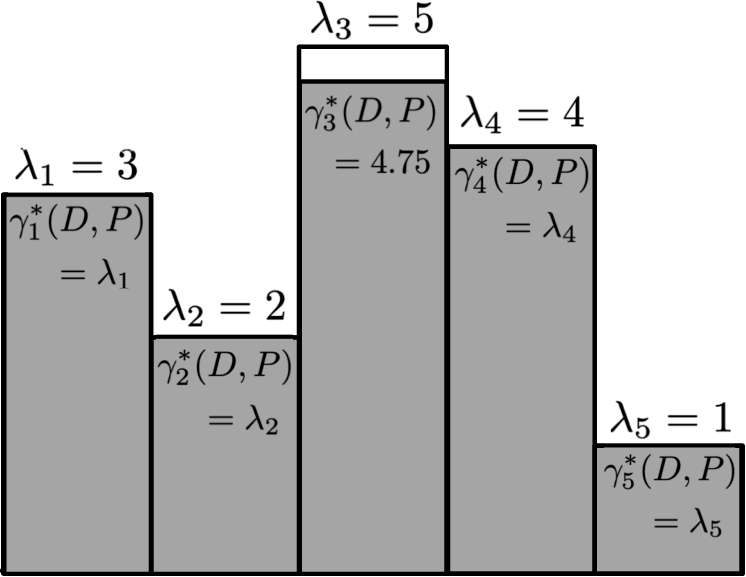}
  \label{fig:P0Dlargea}
}
\hspace{2cm}
\subfigure[Low distortion; no perception constraint] 
{
  \includegraphics[width=0.3\textwidth]{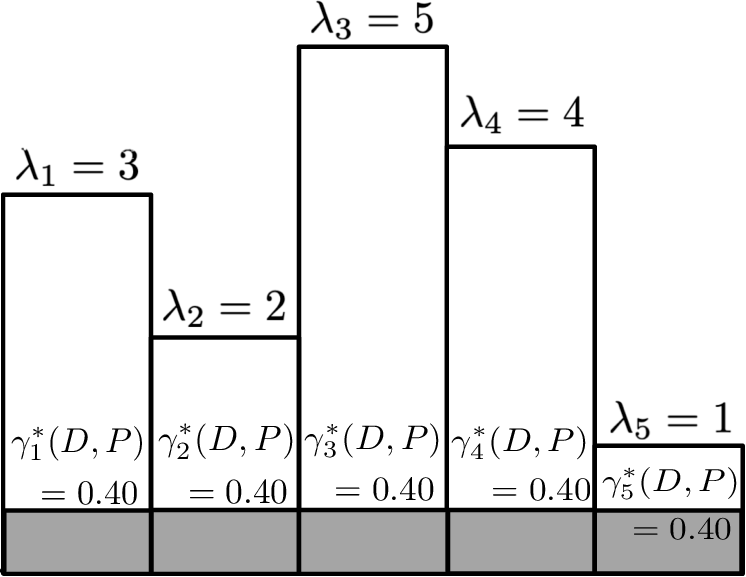}
  \label{fig:P0Dsmallb}
}
\subfigure[High distortion; zero perception loss] 
{
	\includegraphics[width=0.3\textwidth]{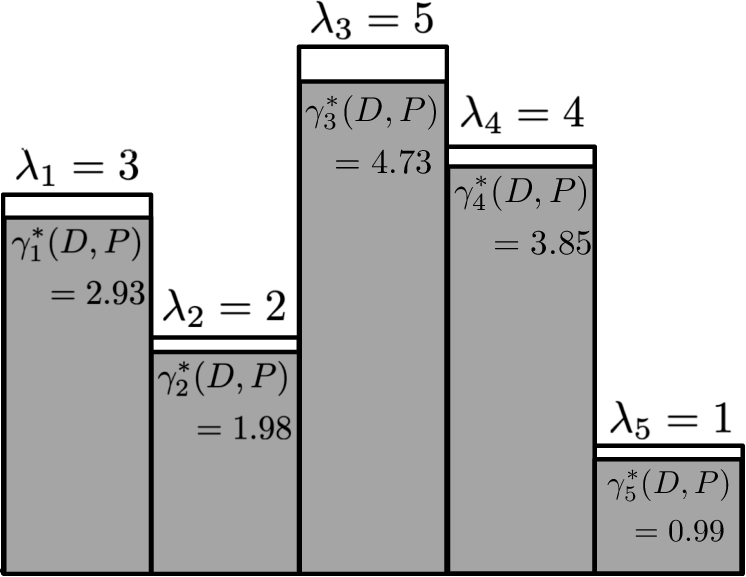}
	\label{fig:P0Dlargec}
}
\hspace{2cm}
\subfigure[Low distortion; zero perception loss] 
{
	\includegraphics[width=0.3\textwidth]{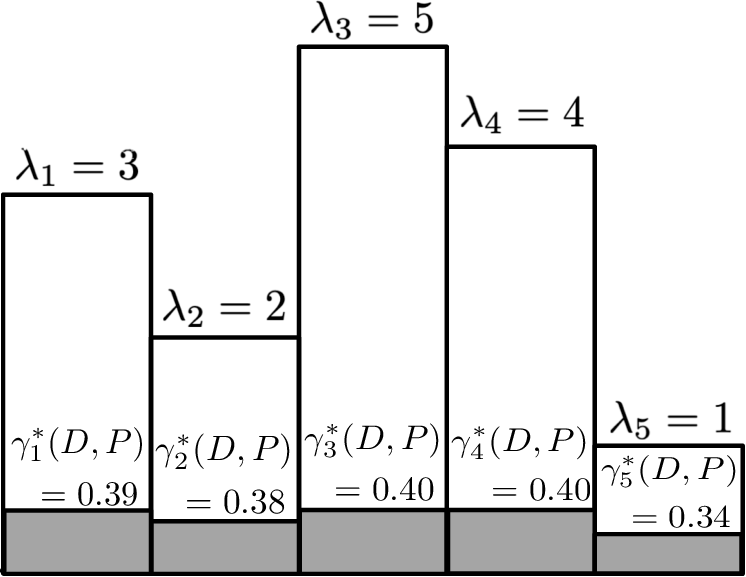}
	\label{fig:P0Dsmalld}
}
\caption{The water levels assigned to different components for a Gaussian vector source
with $\lambda_1=3, \lambda_2=2, \lambda_3=5, \lambda_4=4$ and $\lambda_5=1$.}
\label{fig:water-level}
\end{figure*}

Fig.~\ref{fig:water-level} shows the water levels of different components for
both low-distortion and high-distortion compression with $P=\infty$  or
$P=0$ for an example of a Gaussian vector source. The water levels determine 
the compression rates assigned to each component. 

In Fig.~\ref{fig:water-level}(a), for high-distortion compression with no perception constraint, all components except the one with the largest eigenvalue are allocated a zero
compression rate (cf.\ Proposition~\ref{low-rate-large-P}). With an
active perception constraint, as shown in Fig.~\ref{fig:water-level}(c) for the
$P=0$ case, all components are allocated positive rates (cf.\ Proposition~\ref{low-rate-0-PLF}). 

In Fig.~\ref{fig:water-level}(b), for low-distortion compression with no perception constraint, the water levels of all components are the same (cf.\ Proposition~\ref{high-rate-large-P}). 
At low distortion and with an active perception constraint,  as shown in Fig.~\ref{fig:water-level}(d) for the $P=0$ case, the water levels of different
components are approximately equal with some slight differences which are determined
by \eqref{low-D-P0-water-level} in Proposition~\ref{high-rate-0-PLF}.
Therefore, in the low-distortion regime, the water levels of all components are approximately the same regardless of the perception constraint.

\section{Conclusions}

This paper characterizes the RDP function for a Gaussian vector source. In contrast to the traditional reverse water-filling solution (without a perception constraint), the water levels assigned to different components are not necessarily equal. 
When both distortion and perception constraints are active, every component is assigned a positive rate. These results have implications to perception-aware image coding. 


\appendices

\section{Asymptotic Analysis of the Traditional RD Function}

\label{asymptotic-RD-app}

\subsection{High-Distortion Compression}
\label{low-asym}
Here, we consider $D=\sum_{\ell=1}^L\lambda_{\ell}-\epsilon$ for sufficiently small $\epsilon>0$. 
Without loss of generality, we assume that the eigenvalues are ordered as follows 
\begin{IEEEeqnarray}{rCl}
	\lambda_1\leq \lambda_2\leq \ldots\leq\lambda_{L}.
\end{IEEEeqnarray}
First consider the case that $|\mathcal{L}^{\max}|=1$. The distortion constraint~\eqref{distortion-budget} implies that 
\begin{IEEEeqnarray}{rCl}
	\sum_{\ell=1}^L [\lambda_{\ell}-\nu(D)]^+=\epsilon.\label{D-high}
\end{IEEEeqnarray}
The above condition implies that for a small enough $\epsilon>0$, $\nu(D)$ should satisfy 
\begin{IEEEeqnarray}{rCl}
	\lambda_{1}\leq \lambda_2\leq \ldots\leq \lambda_{L-1}\leq \nu(D)< \lambda_{L}. \label{water-level-condition}
\end{IEEEeqnarray}
Considering~\eqref{water-level-condition} with~\eqref{D-high} yields 
\begin{IEEEeqnarray}{rCl}
	\nu(D)=\lambda_L-\epsilon.
\end{IEEEeqnarray}
Plugging the above into the RDP function of Proposition~\ref{traditional-RD}, we get
\begin{IEEEeqnarray}{rCl}
	R\left(\sum_{\ell=1}^L\lambda_{\ell}-\epsilon,\infty\right)&=&\frac{1}{2}\log \frac{\lambda_L}{\lambda_L-\epsilon}\\
	&=& \frac{1}{2\lambda_L}\epsilon+O(\epsilon^2).
\end{IEEEeqnarray}
Finally, noting $\lambda_L=\max_{\ell}\lambda_{\ell}$ gives \eqref{eq:RD_rate_high_distortion}.

If $|\mathcal{L}^{\max}|>1$, then similar to the above discussion, all eigenvalues except the largest ones are assigned a zero compression rate and for the maximum eigenvalues, we have the following water level 
\begin{IEEEeqnarray}{rCl}
	\nu(D)=\lambda^{\max}-\frac{\epsilon}{|\mathcal{L}^{\max}|},
\end{IEEEeqnarray}
and the following rate
\begin{IEEEeqnarray}{rCl}
	R\left(\sum_{\ell=1}^L\lambda_{\ell}-\epsilon,\infty\right)&=&\frac{|\mathcal{L}^{\max}|}{2}\log \frac{\lambda_L}{\lambda_L-\frac{\epsilon}{|\mathcal{L}^{\max}|}}\\
	&=& \frac{1}{2\lambda_L}\epsilon+O(\epsilon^2).
\end{IEEEeqnarray}
This proves \eqref{eq:RD_rate_high_distortion} for arbitrary $\mathcal{L}^{\max}$. 

\subsection{Low-Distortion Compression}
\label{high-asym}

Consider the case of $D=\epsilon$ for sufficiently small $\epsilon>0$.
In this low-distortion regime, the constant water level $\nu(D)$ is not saturated by the eigenvalues. Thus, Proposition~\ref{traditional-RD} simplifies to the following
\begin{IEEEeqnarray}{rCl}
	R(\epsilon,\infty)=\frac{1}{2}\sum_{\ell=1}^L\log \frac{\lambda_{\ell}}{\nu(D)}.\label{R-large}
\end{IEEEeqnarray}
Also, the distortion constraint \eqref{distortion-budget} implies that 
\begin{IEEEeqnarray}{rCl}
	\nu(D)=\frac{D}{L}.\label{D-large}
\end{IEEEeqnarray}
Combining \eqref{R-large} and \eqref{D-large}, we get the rate expression \eqref{eq:RD_rate_low_distortion} in Proposition \ref{high-rate-large-P}.

\section{Proof of Theorem \ref{thm-Gaussian-opt}}\label{thm-Gaussian-opt-proof}

First, we prove the optimality of Gaussian reconstruction for the case of the KL-divergence as the perception metric. Define the following distribution
	\begin{IEEEeqnarray}{rCl}
		P_{\hat{X}^*|X}=\arg \min_{\substack{P_{\hat{X}|X}:\\ \mathbb{E}[\|X-\hat{X}\|^2]\leq D\\D(P_{\hat{X}}\|P_X)\leq P}} I(X;\hat{X}).
	\end{IEEEeqnarray}
 Now, let $\hat{X}_G$ be a random variable jointly Gaussian distributed with $X$ such that 
	\begin{subequations}\label{G-rv-cons}
		\begin{IEEEeqnarray}{rCl}
			\mathbbm{E}[\hat{X}_G]&=&\mathbbm{E}[\hat{X}^*],\\
			\text{cov}(\hat{X}_G,X)&=&\text{cov}(\hat{X}^*,X).
		\end{IEEEeqnarray}
	\end{subequations}
	We proceed with lower bounding the rate  as follows
	\begin{IEEEeqnarray}{rCl}
		I(X;\hat{X}^*)&=& h(X)-h(X|\hat{X}^*)\\
		&\geq & h(X)-h(X|\hat{X}_G)\label{diff-entropy-step}\\
		&=& I(X;\hat{X}_G),\label{Gaussian-codebook-lower-bound-rate}
	\end{IEEEeqnarray}
	where~\eqref{diff-entropy-step} follows from~\eqref{G-rv-cons} and the fact that under a fixed covariance matrix, a jointly Gaussian distribution maximizes the conditional differential entropy \cite[Lemma 2]{Jun-differential-entropy}. The condition~\eqref{G-rv-cons} also implies that for the distortion loss, we have
	\begin{IEEEeqnarray}{rCl}
		D\geq \mathbbm{E}[\|X-\hat{X}^*\|^2]=\mathbbm{E}[\|X-\hat{X}_G\|^2].\;\;\label{Gaussian-codebook-lower-bound-distortion}
	\end{IEEEeqnarray}
	Moreover, for the perception loss, we have
	\begin{IEEEeqnarray}{rCl}
		D(P_{\hat{X}^*}\|P_X) &=& \int P_{\hat{X}^*}(x)\log \frac{P_{\hat{X}^*}(x)}{P_{X}(x)}dx\\
		&=& -h(\hat{X}^*)-\int P_{\hat{X}^*}(x)\log P_{X}(x)dx\\
		&=& -h(\hat{X}^*)+\frac{1}{2}\int P_{\hat{X}^*}(x) x\Sigma_X^{-1}x^Tdx\nonumber\\&&\hspace{0.5cm}+\frac{1}{2}\log (2\pi)^L \det(\Sigma_X)\\
		&=&-h(\hat{X}^*)+\frac{1}{2}\int P_{\hat{X}_G}(x) x\Sigma_X^{-1}x^Tdx\nonumber\\&&\hspace{0.5cm}+\frac{1}{2}\log (2\pi)^L \det(\Sigma_X)\label{ineq-kl1}\\
		&=& -h(\hat{X}^*)-\int P_{\hat{X}_G}(x)\log P_{X}(x)dx\\
		&\geq & -h(\hat{X}_G)-\int P_{\hat{X}_G}(x)\log P_{X}(x)dx\label{ineq-kl2}\\
		&=& D(P_{\hat{X}_G}\|P_X),
\end{IEEEeqnarray}
where \eqref{ineq-kl1} follows because the expression $x\Sigma_X^{-1}x^T$ for a vector $x=(x_1,\ldots,x_L)$ only contains the terms such as $x_{\ell}^2$, $x_{\ell}$ and $x_{\ell}x_{\ell'}$ for $\ell,\ell'\in\{1,\ldots,L\}$, and since according to~\eqref{G-rv-cons},  $\hat{X}^*$ has the same mean and covariance matrix as $\hat{X}_G$, the expected values of these terms with respect to $P_{\hat{X}^*}$ are equal to the same expectations calculated with respect to $P_{\hat{X}_G}$; \eqref{ineq-kl2} follows because for a fixed covariance matrix, the differential entropy is maximized by a Gaussian distribution \cite[Thm 8.6.5]{Cover}. Finally, there is no loss of optimality in setting $\mathbb{E}[\hat{X}_G]=0$ since replacing  $\hat{X}_G$ with $\hat{X}_G-\mathbb{E}[\hat{X}_G]$ does not increase $I(X;\hat{X}_G)$, $\mathbb{E}[\|X-\hat{X}_G\|^2]$, and $D(P_{\hat{X}_G}\|P_X)$.

 Thus, replacing $\hat{X}^*$ by $\hat{X}_G$ does not increase the rate, while distortion and perception constraints remain satisfied. Thus, the optimal $\hat{X}^*$ must be jointly Gaussian with $X$.

For the case of the Wasserstein-2 distance as the perception metric,   lower bounding steps for $I(X;\hat{X}^*)$ and $\mathbbm{E}[\|X-\hat{X}^*\|^2]$ are the same as~\eqref{Gaussian-codebook-lower-bound-rate} and~\eqref{Gaussian-codebook-lower-bound-distortion}, respectively. For the perception metric, the steps are refined as follows. Define the following distribution
\begin{IEEEeqnarray}{rCl}
	P_{U^*V^*}= \arg\inf_{\substack{\tilde{P}_{UV}:\\\tilde{P}_U=P_{X}\\\tilde{P}_{V}=P_{\hat{X}^*}}}\mathbbm{E}_{\tilde{P}}[\|U-V\|^2].\label{step4}
\end{IEEEeqnarray}
Let $P_{U_GV_G}$ be a joint Gaussian distribution such that
\begin{subequations}\label{covariance-G-match}
		\begin{IEEEeqnarray}{rCl}
			\mathbbm{E}[U_G]&=&\mathbbm{E}[U^*],\\
   \mathbbm{E}[V_G]&=&\mathbbm{E}[V^*],\\
			\text{cov}(U_G,V_G)&=&\text{cov}(U^*,V^*).
		\end{IEEEeqnarray}
	\end{subequations}
Then, we have the following set of inequalities: 
\begin{IEEEeqnarray}{rCl}
	P&\geq& W_2^2(P_{X},P_{\hat{X}^*}) \\
	&=& \inf_{\substack{\tilde{P}_{UV}:\\\tilde{P}_U=P_{X}\\\tilde{P}_{V}=P_{\hat{X}^*}}}\mathbbm{E}_{\tilde{P}}[\|U-V\|^2]\label{step5}\\
	&=& \mathbbm{E}[\|U^*-V^*\|^2]\label{P-step}\\
	&=& \mathbbm{E}[\|U_G-V_G\|^2]\label{step6}\\
	&\geq &  W_2^2(P_{U_G},P_{V_G})\\
	&=& \inf_{\substack{\hat{P}_{UV}:\\\hat{P}_U=P_{U_G}\\\hat{P}_V=P_{V_G}}}\mathbbm{E}_{\hat{P}}[\|U-V\|^2]\\
	&=& \inf_{\substack{\hat{P}_{UV}:\\\hat{P}_U=P_{X}\\\hat{P}_V=P_{\hat{X}_G}}}\mathbbm{E}_{\hat{P}}[\|U-V\|^2]\label{step7}\\
	&=& W_2^2(P_{X},P_{\hat{X}_G}),\label{step8}
\end{IEEEeqnarray}
where
\begin{itemize}
	\item \eqref{P-step} follows from the definition in \eqref{step4};
	\item \eqref{step6} follows from~\eqref{covariance-G-match} which states that $(U^*,V^*)$ and $(U^G,V^G)$ have the same first- and second-order statistics;
	\item \eqref{step7} follows because $P_{V_G}=P_{\hat{X}_G}$ and $P_{U_G}=P_{X}$,
		which are justified as follows. First, notice that both $P_{V_G}$ and $P_{\hat{X}_G}$ are Gaussian distributions.  According to~\eqref{covariance-G-match}, the first- and second-order statistics of $V_G$ are equal to those of $V^*$. Also, from~\eqref{step4}, we know that $P_{V^*}=P_{\hat{X}^*}$, hence the first- and second-order statistics  of $V^*$ and $\hat{X}^*$ are the same. On the other side, from~\eqref{G-rv-cons}, we know that the first- and second-order statistics of $\hat{X}^*$ are equal to those of $\hat{X}_G$. Thus, we conclude that  $P_{V_G}=P_{\hat{X}_G}$. A similar argument shows that $P_{U_G}=P_{X}$.
\end{itemize}
 Thus, without loss of optimality one can replace $\hat{X}^*$ by $\hat{X}_G$ since the rate does not increase, while the distortion and perception constraints remain satisfied.


\section{Proof of Theorem~\ref{KL-thm}}\label{KL-app}

We aim to establish the RDP function for the case of KL-divergence as the perception metric by showing that
\begin{IEEEeqnarray}{rCl}
R(D,P)=R^*(D,P),\label{lower-bound-RDP}
\end{IEEEeqnarray}
where
\begin{subequations}\label{Rstar}
\begin{IEEEeqnarray}{rCl}
	&&\hspace{-0.5cm}R^*(D,P)\nonumber\\
	&&=\min_{\{\hat{\lambda}_{\ell},\gamma_{{\ell}}\}_{\ell=1}^L} \frac{1}{2}\sum_{\ell=1}^L \log \frac{\lambda_{\ell}}{\gamma_{{\ell}}}\label{RpDP_obj}\\
	&&\hspace{0.8cm}\text{s.t.}\hspace{0.6cm} 0<\gamma_{\ell}\leq \lambda_{\ell} \label{eq:gamma_constraint}\\
	&&\hspace{1.8cm} 0  \leq \hat{\lambda}_{\ell} \label{eq:lambda_constraint}\\
	&&\hspace{1.8cm}\sum_{\ell=1}^L
	\left( \lambda_{{\ell}}-2\sqrt{\hat{\lambda}_{{\ell}}(\lambda_{{\ell}}-\gamma_{{\ell}})}+\hat{\lambda}_{{\ell}} \right) \leq D \;\;\;\;\;\;\label{cons1-program}\\
	&&\hspace{1.8cm}\frac{1}{2}\sum_{\ell=1}^L \left(\frac{\hat{\lambda}_{{\ell}}}{\lambda_{{\ell}}}-1+\log\frac{\lambda_{{\ell}}}{\hat{\lambda}_{{\ell}}}\right)\leq P.\label{RpDP_P_cons}
\end{IEEEeqnarray}
\end{subequations}

\subsection{Proof of $R^*(D,P)\geq R(D,P)$}
 Let $\{\gamma_{\ell},\hat{\lambda}_{\ell}\}_{\ell=1}^L$
 be the optimal solution of~\eqref{Rstar}. For $\ell\in\{1,\ldots,L\}$, let $\hat{Z}_{G,\ell}^*$ be jointly Gaussian with $Z_{\ell}$ with their covariance matrix as given in~\eqref{eq:covariance_Z_hat}, and be independent of all other $Z_{\ell'}$, i.e., $\forall \ell' \neq \ell$. Let $\hat{Z}_{G}^*=(\hat{Z}_{G,1}^*,\ldots, \hat{Z}_{G,L}^*)$. Further, set $\hat{X}^*_G=\Theta^T \hat{Z}^*_G$.
It can be verified that
\begin{IEEEeqnarray}{rCl}
\mathbb{E}[\|X-\hat{X}^*_G\|^2]&=&\mathbb{E}[\|Z-\hat{Z}^*_G\|^2]\label{eq:invariance1}\\
&=&\sum\limits_{\ell=1}^L\mathbb{E}[(Z_{\ell}-\hat{Z}^*_{G,\ell})^2]\\
&=&\sum_{\ell=1}^L
	\left( \lambda_{{\ell}}-2\sqrt{\hat{\lambda}_{{\ell}}(\lambda_{{\ell}}-\gamma_{{\ell}})}+\hat{\lambda}_{{\ell}}
	\right) \;\;\\&\leq& D,
\end{IEEEeqnarray}
and
\begin{IEEEeqnarray}{rCl}
D(P_{X^*_{G}}\|P_X)&=&D(P_{\hat{Z}^*_{G}}\|P_Z)\label{eq:invariance2}\\
&=&\sum\limits_{\ell=1}^LD(P_{\hat{Z}^*_{G,\ell}}\|P_{Z_{\ell}})\\
&=&\frac{1}{2}\sum_{\ell=1}^L \left(\frac{\hat{\lambda}_{{\ell}}}{\lambda_{{\ell}}}-1+\log\frac{\lambda_{{\ell}}}{\hat{\lambda}_{{\ell}}}\right)\\
&\leq& P,
\end{IEEEeqnarray}
where \eqref{eq:invariance1} and \eqref{eq:invariance2} are due to the invariance of KL-divergence and Euclidean distance under unitary transformations. 
Therefore, we must have $R(D,P)\leq I(X;\hat{X}^*_G)$. On the other hand,
\begin{IEEEeqnarray}{rCl}
I(X;\hat{X}^*_G)&=&I(Z;\hat{Z}^*_G)\\
&=&\sum\limits_{\ell=1}^LI(Z_{\ell};\hat{Z}^*_{G,\ell})\\
&=&\frac{1}{2}\sum\limits_{\ell=1}^L\log\frac{\lambda_{\ell}}{\gamma_{\ell}}\\
&=&R^*(D,P).
\end{IEEEeqnarray}
This proves $R^*(D,P)\geq R(D,P)$.

\subsection{Proof of $R^*(D,P)\leq R(D,P)$}

It follows from Theorem \ref{thm-Gaussian-opt} that
\begin{subequations}\label{RDP-recall}
\begin{IEEEeqnarray}{rCl}
	R(D,P)=& \inf_{P_{\hat{X}_G|X}} & I(X;\hat{X}_G) \\
	 & \text{s.t.}& \mathbb{E}[\|X-\hat{X}_G \|^2]\leq D  \\
	 & & D(P_{\hat{X}_G}\|P_{X})\leq P,
\end{IEEEeqnarray}
\end{subequations}
where $\hat{X}_G$ has mean zero and is jointly Gaussian with $X$.
Let $P_{\hat{X}_G^*|X}$ be the optimal distribution of the program in~\eqref{RDP-recall} and define $\hat{Z}^*_G=\Theta \hat{X}^*_G$.  Let $\Sigma_{\hat{X}^*_G}$ be the covariance matrix of $\hat{X}^*_G$ and $\Lambda_{\hat{Z}^*_G}$ be a diagonal matrix whose diagonal elements coincide with those of $\Theta \Sigma_{\hat{X}^*_G}\Theta^T$, i.e.,
\begin{IEEEeqnarray}{rCl}
\Lambda_{\hat{Z}^*_G}= \text{diag}^L(\hat{\lambda}_{1},\ldots,\hat{\lambda}_{L}).\label{LambdaZhatl}
\end{IEEEeqnarray}
 Furthermore, define
\begin{equation}
\gamma_{\ell}=\mathbbm{E}[(Z_{\ell}-\mathbb{E}[Z_{\ell}|\hat{Z}^*_{G,\ell}])^2], \;\;\; \ell\in\{1,\ldots,L\}.
\end{equation}
Clearly, \eqref{eq:gamma_constraint} and \eqref{eq:lambda_constraint} are satisfied.

It can be verified that
\begin{IEEEeqnarray}{rCl}
		I(X;\hat{X}^*_G) &=& I(Z;\hat{Z}^*_G)\label{eq1}\\
		&=& h(Z)-h(Z|\hat{Z}^*_G)\\
		&=& \sum_{\ell=1}^Lh(Z_{\ell})-h(Z|\hat{Z}^*_G)\label{eq2}\\
		&\geq &\sum_{\ell=1}^Lh(Z_{\ell})-\sum_{\ell=1}^Lh(Z_{\ell}|\hat{Z}^*_{G,\ell})\label{eq3}\\
		&= &\sum_{\ell=1}^Lh(Z_{\ell})-\sum_{\ell=1}^Lh(Z_{\ell}-\mathbbm{E}[Z_{\ell}|\hat{Z}^*_{G,\ell}]|\hat{Z}^*_{G,\ell})\nonumber\\\\
		&= &\sum_{\ell=1}^Lh(Z_{\ell})-\sum_{\ell=1}^Lh(Z_{\ell}-\mathbbm{E}[Z_{\ell}|\hat{Z}^*_{G,\ell}])\label{eq4a}\\
		&=& \sum_{\ell=1}^L \frac{1}{2}\log \left((2\pi e) \lambda_{{\ell}}\right)-\sum_{\ell=1}^L\frac{1}{2}\log \left((2\pi e)\gamma_{{\ell}}\right)\nonumber\\\label{eq4}\\
		&=& \sum_{\ell=1}^L\frac{1}{2}\log \frac{\lambda_{{\ell}}}{\gamma_{{\ell}}},\label{R-final-expression}
\end{IEEEeqnarray}
where 
\begin{itemize}
	\item \eqref{eq1} is due to the invertibility of unitary transformations;
	\item \eqref{eq2} follows because $Z_1,\ldots, Z_{L}$ are independent;
	\item \eqref{eq3} follows from the chain rule and that conditioning does not increase entropy;
	\item \eqref{eq4a} follows because $Z_{\ell}-\mathbbm{E}[Z_{\ell}|\hat{Z}^*_{G,\ell}]$
		is independent of $\hat{Z}^*_{G,\ell}$; 
	\item \eqref{eq4} follows because $\mathbb{E}[Z_{\ell}^2]=\lambda_{{\ell}}$ and  
			$\mathbb{E}[(Z_{\ell}-\mathbb{E}[Z_{\ell}|\hat{Z}^*_{G,\ell}])^2]=\gamma_{{\ell}}$.
\end{itemize}

Next, consider the expected distortion loss as follows:
\begin{IEEEeqnarray}{rCl}
	D&\geq& \mathbb{E}[\| X-\hat{X}^*_G\|^2] \\
	&=&\mathbb{E}[\| Z-\hat{Z}^*_G\|^2]\label{eq5}\\
	&=&\sum_{\ell=1}^L \mathbb{E}[ (Z_{\ell}-\hat{Z}^*_{G,\ell})^2]\\
	&=&\sum_{\ell=1}^L \mathbb{E}[ Z_{\ell}^2]-2\mathbb{E}[Z_{\ell}\hat{Z}^*_{G,\ell}]+\mathbb{E}[(\hat{Z}^*_{G,\ell})^2] \\
	&=& \sum_{\ell=1}^L \lambda_{{\ell}}-2\mathbb{E}[Z_{\ell}\hat{Z}^*_{G,\ell}]+\hat{\lambda}_{{\ell}}\label{eq10}\\
	&=& \sum_{\ell=1}^L \lambda_{{\ell}}-2\sqrt{\hat{\lambda}_{{\ell}}(\lambda_{{\ell}}-\gamma_{{\ell}})}+\hat{\lambda}_{{\ell}},\label{eq11}
\end{IEEEeqnarray}
where 
\begin{itemize}
	\item \eqref{eq5} is due to
 the invariance of Euclidean distance under unitary transformations,
	\item \eqref{eq10} follows because $\mathbb{E}[Z_{\ell}^2]=\lambda_{{\ell}}$ and $\mathbb{E}[(\hat{Z}^*_{G,\ell})^2]=\hat{\lambda}_{{\ell}}$,
	\item \eqref{eq11} follows from the identity 
		\begin{equation}
		\mathbb{E}[(Z_{\ell}-\mathbb{E}[Z_{\ell}|\hat{Z}^*_{G,\ell}])^2]=\mathbb{E}[Z_{\ell}^2]-(\mathbb{E}[Z_{\ell}\hat{Z}^*_{G,\ell}])^2(\mathbb{E}[\hat{Z}^*_{G,\ell}])^{-1}
		\end{equation}
		and
		\begin{IEEEeqnarray}{rCl}
		\mathbb{E}[(Z_{\ell}-\mathbb{E}[Z_{\ell}|\hat{Z}^*_{G,\ell}])^2] & = & \gamma_{\ell}, \\
		\mathbb{E}[Z_{\ell}^2] & = & \lambda_{{\ell}}, \\ 
		\mathbb{E}[(\hat{Z}^*_{G,\ell})^2] & = & \hat{\lambda}_{{\ell}}.
		\end{IEEEeqnarray}
\end{itemize}

Finally, consider the perception loss:
\begin{IEEEeqnarray}{rCl}
	P&\geq& D(P_{\hat{X}^*_G}\|P_X) \\&=& \frac{1}{2}\Bigg(\text{tr}(\Lambda_X^{-1}\Theta\Sigma_{\hat{X}^*_G}\Theta^T)-L+\log \frac{\det(\Lambda_X)}{\det(\Theta \Sigma_{\hat{X}^*_G}\Theta^T)}\Bigg)\nonumber\\\\
	&=&\frac{1}{2}\left(\text{tr}(\Lambda_X^{-1}\Lambda_{\hat{Z}^*_G})-L+\log \frac{\det(\Lambda_X)}{\det(\Theta \Sigma_{\hat{X}^*_G}\Theta^T)}\right)\label{eq6}\\
	&\geq & \frac{1}{2}\left(\text{tr}(\Lambda_X^{-1}\Lambda_{\hat{Z}^*_G})-L+\log \frac{\det(\Lambda_X)}{\det(\Lambda_{\hat{Z}^*_G})}\right)\label{eq7}\\
	&=& \frac{1}{2}\sum_{\ell=1}^L \left(\frac{\hat{\lambda}_{{\ell}}}{\lambda_{{\ell}}}-1+\log\frac{\lambda_{{\ell}}}{\hat{\lambda}_{{\ell}}}\right),\label{perc-simplified}
\end{IEEEeqnarray}
where                \begin{itemize}
	\item \eqref{eq6} follows because $\Lambda_X^{-1}$ is a diagonal matrix and thus the trace depends on the diagonal elements of $\Theta\Sigma_{\hat{X}^*_G}\Theta^T$ which are equal to the diagonal elements of $\Lambda_{\hat{Z}^*_G}$,
	\item \eqref{eq7} follows from the Hadamard's inequality  for a positive semidefinite matrix.
\end{itemize}
Combining~\eqref{R-final-expression},~\eqref{eq11}, and~\eqref{perc-simplified} yields $R^*(D,P)\leq R(D,P)$.


\section{Proof of Theorem~\ref{thm-optimal-sol}}\label{thm-optimal-sol-proof}

First, we show that 
the optimization problem in~\eqref{Rstar} is convex. 
The second derivative of the objective function \eqref{RpDP_obj} with respect to $\gamma_{\ell}$ is $\frac{1}{2\gamma_{\ell}^2}$ which is positive. The second derivative of the function in the constraint \eqref{RpDP_P_cons} with respect to $\hat{\lambda}_{\ell}$ is $\frac{1}{2\hat{\lambda}_{\ell}^2}$ which is again positive. 

It remains to study the constraint~\eqref{cons1-program}. The Hessian matrix of the function in this constraint is 
\begin{IEEEeqnarray}{rCl}
	\begin{bmatrix} \frac{\sqrt{\lambda_{\ell}-\gamma_{\ell}}}{2\sqrt{\hat{\lambda}^3_{\ell}}}&\frac{1}{2\sqrt{\hat{\lambda}_{\ell}(\lambda_{\ell}-\gamma_{\ell})}}\\ \frac{1}{2\sqrt{\hat{\lambda}_{\ell}(\lambda_{\ell}-\gamma_{\ell})}}&\frac{\sqrt{\hat{\lambda}_{\ell}}}{2\sqrt{(\lambda_{\ell}-\gamma_{\ell})^3}}\end{bmatrix}.
\end{IEEEeqnarray}
The determinant of the above matrix is zero, and the matrix has positive diagonal terms. Thus, it is a positive semidefinite matrix, which implies the convexity of the associated function.
This proves the convexity of the program in~\eqref{Rstar}. 

Since $(D,P)$ is assumed to be strictly feasible, the Slater's condition is satisfied.
This implies that the solution to this problem is equal to that of the following 
dual optimization problem
\begin{IEEEeqnarray}{rCl}
	&&\max_{\nu_1, \nu_2, \eta_\ell, \xi_\ell \ge 0}\;\; \min_{\{\gamma_{\ell},\hat{\lambda}_{\ell}\}_{\ell=1}^L}  \;\;\frac{1}{2}\sum_{\ell=1}^L \log \frac{\lambda_{\ell}}{\gamma_{{\ell}}}\nonumber\\&&\hspace{1cm}+\nu_1 \left(\sum_{\ell=1}^L
	( \lambda_{{\ell}}-2\sqrt{\hat{\lambda}_{{\ell}}(\lambda_{{\ell}}-\gamma_{{\ell}})}+\hat{\lambda}_{{\ell}}
	) -D\right)\nonumber\\
&&\hspace{1cm} +\nu_2 \left(\frac{1}{2}\sum_{\ell=1}^L \left(\frac{\hat{\lambda}_{{\ell}}}{\lambda_{{\ell}}}-1+\log\frac{\lambda_{{\ell}}}{\hat{\lambda}_{{\ell}}}\right)- P\right)\nonumber\\&&\hspace{1cm}+\sum_{\ell=1}^L\xi_{\ell}(\gamma_{\ell}-\lambda_{\ell})-\sum_{\ell=1}^L\eta_{\ell}\hat{\lambda}_{\ell}, \label{Lagrange-dual-function}
\end{IEEEeqnarray}
where $\{\nu_{1},\nu_2\}$ and $\{\xi_{\ell},\eta_{\ell}\}_{\ell=1}^L$ are nonnegative Lagrange multipliers.
Note that the distortion function has implicit constraints $\hat\lambda_\ell \ge 0$
and $\gamma_\ell \le \lambda_\ell $. Moreover, the derivatives of the respective terms 
go to infinity when $\hat\lambda_\ell$ and $\gamma_\ell$ approach these boundaries. 
For this reason, we cannot immediately write down the Karush-Kuhn-Tucker (KKT) conditions for the optimization problem, and instead, need to carefully consider the behaviour of 
the optimization problem close to these boundaries. 
Toward this end, we consider the following three different cases.
%
%

\subsection{Maximum of Outer Optimization Occurs at $\nu_1,\nu_2>0$} 

This is the case where both perception and distortion constraints are active. 
Let $\hat{\lambda}^*_{\ell}$ and $\gamma_\ell^*$ be the optimal solution to
the inner minimization problem in \eqref{Lagrange-dual-function} for 
the optimal $\nu_1$ and $\nu_2$.  

We first note that
\begin{IEEEeqnarray}{rCl}
\hat{\lambda}^*_{\ell}&>&0.\label{lambda-hat-positive}
\end{IEEEeqnarray}
This is because if $\hat{\lambda}^*_{\ell}=0$, then we have $P=\infty$ which would violate the perception constraint. 

Next, we prove that the following strict inequality holds:
 	\begin{IEEEeqnarray}{rCl}
 	\gamma_{\ell}^*&<&\lambda_{\ell}. \label{gamma-positive}
 	\end{IEEEeqnarray}
We proceed with proof by contradiction.
Suppose that the above strict inequality does not hold, i.e., $\gamma_{\ell}^*=\lambda_{\ell}$. We show that such $\gamma_\ell^*$ cannot be the optimal solution to the inner minimization problem.


The Lagrangian term in \eqref{Lagrange-dual-function} depends on $\gamma_{\ell}$ and $\hat{\lambda}_{\ell}$ through the following function: 
\begin{IEEEeqnarray}{rCl}
	G_{\ell}(\gamma_{\ell},\hat{\lambda}_{\ell})&=&\frac{1}{2}\log\frac{\lambda_{\ell}}{\gamma_{\ell}}+\nu_1\left(\lambda_{\ell}-2\sqrt{\hat{\lambda}_{\ell}(\lambda_{\ell}-\gamma_{\ell})}+\hat{\lambda}_{\ell}\right)\nonumber\\&&\hspace{0.5cm}+\frac{\nu_2}{2}\left(\frac{\hat{\lambda}_{\ell}}{\lambda_{\ell}}-1+\log \frac{\lambda_{\ell}}{\hat{\lambda}_{\ell}}\right)\nonumber\\&&\hspace{0.5cm}+\xi_{\ell}(\gamma_{\ell}-\lambda_{\ell})-\eta_{\ell}\hat{\lambda}_{\ell}.
	\end{IEEEeqnarray}
Fix $\hat\lambda_\ell = \hat\lambda_\ell^*$. When we deviate from $\gamma_{\ell}^*=\lambda_{\ell}$ to $\gamma_{\ell}'=\lambda_{\ell}-\epsilon$ for some small $\epsilon>0$, the first order change in $G_\ell(\gamma_{\ell},\hat{\lambda}_{\ell}^*)$ can be seen as follows:
\begin{IEEEeqnarray}{rCl}
G_{\ell}(\gamma^*_{\ell},\hat{\lambda}^*_{\ell}) - G_{\ell}(\gamma'_{\ell},\hat{\lambda}^*_{\ell})
&=&\frac{1}{2}\log \frac {\lambda_{\ell}-\epsilon} {\lambda_{\ell}}
+ 2\nu_1 \sqrt{\epsilon\hat{\lambda}^*_{{\ell}}} - \epsilon\xi_{\ell}\nonumber\\\\
	&=& - \frac{\epsilon}{2\lambda_{\ell}} + 2\nu_1 \sqrt{\epsilon\hat{\lambda}^*_{{\ell}}} - \epsilon\xi_{\ell} + O(\epsilon^2)\nonumber\\
\label{Lagrange-dual-step1}\\
	&=& 2\nu_1 \sqrt{\epsilon\hat{\lambda}^*_{{\ell}}}+O(\epsilon), \label{Lagrange-dual-step2}
	\end{IEEEeqnarray}	
where we use the fact that $\log (1-x) = -x + O(x^2)$ for small $x$. 
Thus if $\nu_1 > 0$, since $\hat\lambda_\ell^* > 0$, for sufficiently small $\epsilon > 0$, we can strictly decrease $G_{\ell}(\gamma^*_{\ell},\hat{\lambda}^*_{\ell})$, while satisfying the implicit constraints. This contradicts the assumption that $\gamma^*_{\ell}=\lambda_{\ell}$ is the optimal solution to the inner minimization problem. This proves~\eqref{gamma-positive}, which implies that every component has positive rate.


The strict inequalities in~\eqref{gamma-positive} and~\eqref{lambda-hat-positive} imply that in this case, the optimal solution occurs at the interior of the set 
$\{\hat{\lambda}^*_{\ell} \ge 0 \text{\ and\ } \gamma_{\ell}^* \le \lambda_{\ell} \}$.
This allows us to write down the
KKT conditions for the optimal
primal variables $(\gamma_{\ell}^*,\hat{\lambda}_{{\ell}}^*)$ 
and the optimal dual variables $\{\nu_{1},\nu_2\}$ and $\{\xi_{\ell},\eta_{\ell}\}_{\ell=1}^L$ as follows: 
\begin{subequations}\label{kkt-conditions-original}
	\begin{IEEEeqnarray}{rCl}
		\frac{1}{2\gamma^*_{{\ell}}}-\nu_1\sqrt\frac{{\hat{\lambda}_{{\ell}}^*}}{{\lambda_{{\ell}}-\gamma^*_{{\ell}}}}-\xi_{\ell} & = & 0\qquad\;\;\;\label{cond-opta}\\
		\nu_1\left(-\sqrt{\frac{\lambda_{{\ell}}-\gamma^*_{{\ell}}}{\hat{\lambda}^*_{{\ell}}}}+1\right)+\frac{1}{2}\nu_2\left(\frac{1}{\lambda_{{\ell}}}-\frac{1}{\hat{\lambda}^*_{{\ell}}}\right)-\eta_{\ell} & = & 0\;\;\;\label{cond-optb}\\
		\xi_{\ell}(\gamma^*_{\ell}-\lambda_{\ell}) & = & 0\;\;\;\\
		\eta_{\ell}\hat{\lambda}^*_{\ell} &=& 0\;\;\;\label{cons-d} \\
		\nu_1 \left(
		\sum_{\ell=1}^L\left( \lambda_{{\ell}}-2\sqrt{\hat{\lambda}_{{\ell}}^*(\lambda_{{\ell}}-\gamma_{{\ell}}^*)}+\hat{\lambda}_{{\ell}}^*
		\right) -D\right)  & = & 0\;\;\;\label{cons-e}\\
		\nu_2 \left(\sum_{\ell=1}^L\frac{1}{2} \left(\frac{\hat{\lambda}_{{\ell}}^*}{\lambda_{{\ell}}}-1+\log\frac{\lambda_{{\ell}}}{\hat{\lambda}_{{\ell}}^*}\right)- P\right)  & = & 0\;\;\;\label{cons-f}
	\end{IEEEeqnarray}
\end{subequations}
along with primal and dual feasibility constraints, i.e., $\eta_\ell \ge 0$, $\xi_\ell \ge 0$ and (\ref{constraint1-program-thm})-(\ref{constraint4-program-thm}).


Due to the strict inequalities \eqref{gamma-positive} and~\eqref{lambda-hat-positive},
we have that $\xi_\ell = 0$ and $\eta_\ell=0$. 
Then, from condition~\eqref{cond-opta}, we can write $\hat{\lambda}^*_{\ell}$ as follows:
\begin{IEEEeqnarray}{rCl}
	\hat{\lambda}_{\ell}^*=\frac{\lambda_{\ell}-\hat{\gamma}^*_{\ell}}{4\gamma^{*2}_{\ell}\nu_1^2}.\label{lambdahat-gamma}
\end{IEEEeqnarray}
Plugging~\eqref{lambdahat-gamma} into~\eqref{cond-optb} yields the following second-order equation in $\gamma_{\ell}^*$
\begin{IEEEeqnarray}{rCl}
		\nu_1(1-2\nu_1\gamma^*_{\ell})=\frac{1}{2}\nu_2\left(\frac{4\gamma_{\ell}^{*2}\nu_1^2}{\lambda_{\ell}-\gamma_{\ell}^*}-\frac{1}{\lambda_{\ell}} \right).\;\;\;\;\;\;\label{eq:unique}
	\end{IEEEeqnarray}
	Note that as $\gamma^*_{\ell}$ varies from $0$ to $\lambda_{\ell}$, the left-hand side of \eqref{eq:unique} decreases monotonically from $\nu_1$ to $(1-2\nu_1\lambda_{\ell})\nu_1$ while the right-hand side of \eqref{eq:unique} increases monotonically from $-\frac{\nu_2}{2\lambda_{\ell}}$ to $+\infty$.
	So, this equation has a unique solution in the interval $(0,\lambda_{\ell})$. 
	The equation \eqref{eq:unique} is quadratic, so it can be solved analytically. 
	The solution gives \eqref{eq:gamma_star_KL} and \eqref{eq:lambda_hat_star_KL}.

\subsection{Maximum of Outer Optimization Occurs at $\nu_1>0,\nu_2=0$} 

This is the where the distortion metric is active, but the perception metric is inactive.
Clearly, this reduces to the traditional rate-distortion function.

\subsection{Maximum of Outer Optimization Occurs at $\nu_1=0$} 

This is the case where the distortion metric is inactive, so the inner minimization problem in \eqref{Lagrange-dual-function} decouples into two independent minimizations, one for $\gamma_{\ell}$ and the other one for $\hat{\lambda}_{\ell}$, as shown below 
\begin{IEEEeqnarray}{rCl}
	&&\min_{\{\gamma_{\ell},\hat{\lambda}_{\ell}\}_{\ell=1}^L} \left\{ \frac{1}{2}\sum_{\ell=1}^L \log \frac{\lambda_{\ell}}{\gamma_{{\ell}}} \right. \nonumber \\ 
	&& \qquad \qquad \qquad +\nu_2 \left(\frac{1}{2}\sum_{\ell=1}^L \left(\frac{\hat{\lambda}_{{\ell}}}{\lambda_{{\ell}}}-1+\log\frac{\lambda_{{\ell}}}{\hat{\lambda}_{{\ell}}}\right)- P\right) \nonumber \\
	&& \qquad \qquad \qquad \left. +\sum_{\ell=1}^L\xi_{\ell}(\gamma_{\ell}-\lambda_{\ell})-\sum_{\ell=1}^L\eta_{\ell}\hat{\lambda}_{\ell} \right\} \nonumber\\
	&& = \min_{\{\gamma_{\ell}\}_{\ell=1}^L} \left\{\frac{1}{2}\sum_{\ell=1}^L \log \frac{\lambda_{\ell}}{\gamma_{{\ell}}}+\sum_{\ell=1}^L\xi_{\ell}(\gamma_{\ell}-\lambda_{\ell})\right\} \nonumber \\
	&& \qquad + \min_{\{\hat{\lambda}_{\ell}\}_{\ell=1}^L} \left\{ \nu_2 \Bigg(\frac{1}{2}\sum_{\ell=1}^L \left(\frac{\hat{\lambda}_{{\ell}}}{\lambda_{{\ell}}}-1+\log\frac{\lambda_{{\ell}}}{\hat{\lambda}_{{\ell}}}\right)- P\Bigg) \right. \nonumber \\
	&& \qquad \qquad \qquad \qquad \left. -\sum_{\ell=1}^L\eta_{\ell}\hat{\lambda}_{\ell} \right\}. \label{case-nu10-dual-function}
\end{IEEEeqnarray}

For the first optimization problem in~\eqref{case-nu10-dual-function}, its KKT conditions are given by
\begin{IEEEeqnarray}{rCl}
	\frac{1}{2\gamma^*_{{\ell}}}-\xi_{\ell}&=&0,\label{conda-KKT-xi1}\\
	\xi_{\ell}(\gamma^*_{\ell}-\lambda_{\ell})&=&0.\label{conda-KKT-xi2}
	\end{IEEEeqnarray}
The above two conditions imply that
\begin{IEEEeqnarray}{rCl}
	\gamma_{\ell}^*=\lambda_{\ell}.
	\end{IEEEeqnarray}
So each component has zero rate. 

For the second minimization problem in~\eqref{case-nu10-dual-function}, this is the Lagrangian dual of a feasibility problem with the perception constraint only. 
Thus, we can choose $\hat\lambda_\ell^*$ to satisfy the primal constraints: 
\begin{equation}
	\sum_{\ell=1}^L{P}_{\ell}(\hat{\lambda}^*_{\ell}) \le P, \ \ \text{and} \ \ \ 
	\hat\lambda_\ell^* \ge 0.
	\label{eq:gamma_star_proof}
\end{equation}
Note that despite that the distortion constraint is already assumed to be inactive, we still need to impose an additional distortion constraint
on $\hat\lambda_\ell^*$: 
\begin{equation}
\sum_{\ell=1}^L\lambda_{\ell}+\hat{\lambda}^*_\ell \leq D.
	\label{eq:additional_distortion_proof}
\end{equation}
This is because not all $\hat\lambda_\ell^*$'s satisfying 
\eqref{eq:gamma_star_proof} satisfy the constraint
\eqref{eq:additional_distortion_proof}. 
A constraint being inactive simply means that if the constraint is
removed, there is already at least one optimal solution that 
automatically satisfies the constraint. 
In this case, there are multiple optimal solutions, all giving the 
same objective value (of zero rate). So we need to restrict to the
ones that satisfy \eqref{eq:additional_distortion_proof}.
Note that the left-hand side of \eqref{eq:additional_distortion_proof}
is the distortion of the reconstruction at zero rate.

\section{Proof of Theorem~\ref{W2-thm}}
\label{W2-app}

We now establish the RDP Function with the Wasserstein-2 distance as the perception metric. The proof follows similar steps to those of the KL-divergence metric in Appendix~\ref{KL-app}. We just need to rewrite the lower bounding steps for the perception metric. 
Let $P_{\hat{X}^*_G|X}$ be the optimal conditional distribution of the following optimization program 
\begin{subequations}\label{RDP-recall-W3}
\begin{IEEEeqnarray}{rCl}
	R(D,P)=& \inf_{P_{\hat{X}_G|X}} & I(X;\hat{X}_G) \\
	& \text{s.t.} & \mathbb{E}[\|X-\hat{X}_G \|^2]\leq D  \\ 
	&& W_2^2(P_{X},P_{\hat{X}_G})\leq P,
\end{IEEEeqnarray}
\end{subequations}
where $\hat{X}_G$ has mean zero and is jointly Gaussian with $X$.
Let $\hat{Z}^*_G=\Theta \hat{X}^*_G$ and $\Sigma_{\hat{X}^*_G}$ be the covariance matrix of $\hat{X}^*_G$ and $\Lambda_{\hat{Z}^*_G}$ be a diagonal matrix whose diagonal elements coincide with those of $\Theta \Sigma_{\hat{X}^*_G}\Theta^T$, i.e.,
\begin{IEEEeqnarray}{rCl}
\Lambda_{\hat{Z}^*_G}= \text{diag}^L(\hat{\lambda}_{1},\ldots,\hat{\lambda}_{L}).\label{LambdaZhatl-W2}
\end{IEEEeqnarray}
The lower bounding steps for the perception metric are as follows:
\begin{IEEEeqnarray}{rCl}
	W_2^2(P_{X},P_{\hat{X}^*_G})&=&W_2^2(P_{\Theta X},P_{\Theta\hat{X}^*_G})\label{W2-lower-bounding-step}\\
	&=&W_2^2(P_{Z},P_{\hat{Z}^*_G})\label{def-step4}\\
	&\geq & \sum_{\ell=1}^L W_2^2(P_{Z_{\ell}},P_{\hat{Z}^*_{G,\ell}})\label{single-letter-step5}\\
	&=& \sum_{\ell=1}^L \left(\sqrt{\mathbbm{E}[(Z_{\ell})^2]}-\sqrt{\mathbbm{E}[(\hat{Z}^*_{G,\ell})^2]}\right)^2 \quad\;\;\\
	&=&\sum_{\ell=1}^L \left(\sqrt{\lambda_{{\ell}}}-\sqrt{\hat{\lambda}_{{\ell}}} \right)^2,\label{perception-vector-step2}
\end{IEEEeqnarray}
where
\begin{itemize}
	\item\eqref{W2-lower-bounding-step} follows because $\Theta$ is a unitary matrix and Wasserstein-2 distance is invariant under unitary transformations;
	\item\eqref{def-step4} follows from the definitions $Z=\Theta X$ and $\hat{Z}^*_G=\Theta \hat{X}^*_G$;
	\item\eqref{single-letter-step5} follows from the tensorization property of Wasserstein-2 distance, i.e., for given distributions $P_{X_1X_2}$ and $P_{Y_1Y_2}$, we have $W_2^2(P_{X_1X_2},P_{Y_1Y_2})\geq W_2^2(P_{X_1},P_{Y_1})+W_2^2(P_{X_2},P_{Y_2})$;
	\item\eqref{perception-vector-step2} follows from~\eqref{LambdaZl} and \eqref{LambdaZhatl-W2}.
\end{itemize}
On the other hand, the inequality in~\eqref{single-letter-step5} becomes an equality
if $\hat{X}^*_G=\Theta^T \hat{Z}^*_G$ with $\hat{Z}^*_G$ constructed in such a way that $(Z_{\ell},\hat{Z}^*_{G,\ell})$, $\ell\in\{1,\ldots,L\}$, are mutually independent and their covariance matrices are given by~\eqref{eq:covariance_Z_hat}.
Thus, the RDP function for the Wassertein-2 distance as perception metric is given by the following optimization problem:
\begin{subequations}\label{W-program}
\begin{IEEEeqnarray}{rCl}
&&\hspace{-0.5cm}R(D,P)\nonumber\\&=&\min_{\{\hat{\lambda}_{\ell},\gamma_{{\ell}}\}_{\ell=1}^L} \frac{1}{2}\sum_{\ell=1}^L \log \frac{\lambda_{\ell}}{\gamma_{{\ell}}}\label{obj-func-programb}\\
	&&\hspace{0.5cm}\text{s.t.} \hspace{0.5cm} 0<\gamma_{\ell}\leq \lambda_{\ell} \label{feas1-condb}\\
	&&\hspace{1.4cm} 0 \leq \hat{\lambda}_{\ell} \label{feas2-condb}\\
	&&\hspace{1.4cm}\sum_{\ell=1}^L
	\left( \lambda_{{\ell}}-2\sqrt{\hat{\lambda}_{{\ell}}(\lambda_{{\ell}}-\gamma_{{\ell}})}+\hat{\lambda}_{{\ell}}
	\right) \leq D \qquad\;\;\\
	&&\hspace{1.4cm} \sum_{\ell=1}^L \left(\sqrt{\lambda_{{\ell}}}-\sqrt{\hat{\lambda}_{{\ell}}} \right)^2\leq P.\label{cons2-programb}
	\end{IEEEeqnarray}
\end{subequations}

\section{Proof of Theorem~\ref{cor-W2-optimal-sol}}
\label{cor-W2-optimal-sol-app}


First, note that the optimization problem is convex for the Wasserstein-2 distance as justified below. The argument for the rate and distortion constraints is the same as the KL-divergence metric. The second derivative of the perception constraint in~\eqref{cons2-programb} with respect to $\hat{\lambda}_{\ell}$ is $\frac{1}{2}\sqrt{\frac{\lambda_{\ell}}{\hat{\lambda}^{3}_{\ell}}}$, which is positive. 	
	
The optimization problem can be analyzed in the same way as in Appendix~\ref{thm-optimal-sol-proof}, except the case of $\nu_1,\nu_2>0$, which is discussed as follows. 
Here, we need a different proof to show the inequality 
\begin{IEEEeqnarray}{rCl}
	\hat{\lambda}^*_{\ell}&>&0.\label{lambda-hat-positive-W2}
\end{IEEEeqnarray}
(The proof uses the same technique as the one showing $\gamma_\ell^* < \lambda_\ell$ 
in Appendix~\ref{thm-optimal-sol-proof}-1.) 
Consider the following Lagrange dual optimization \begin{IEEEeqnarray}{rCl}
	&&\max_{\nu_1, \nu_2, \eta_\ell, \xi_\ell \ge 0}\;\; \min_{\{\gamma_{\ell},\hat{\lambda}_{\ell}\}_{\ell=1}^L}  \;\;\frac{1}{2}\sum_{\ell=1}^L \log \frac{\lambda_{\ell}}{\gamma_{{\ell}}}\nonumber\\&&\hspace{1cm}+\nu_1 \left(\sum_{\ell=1}^L
	( \lambda_{{\ell}}-2\sqrt{\hat{\lambda}_{{\ell}}(\lambda_{{\ell}}-\gamma_{{\ell}})}+\hat{\lambda}_{{\ell}}
	) -D\right)\nonumber\\
	&&\hspace{1cm} +\nu_2 \left(\sum_{\ell=1}^L \left(\sqrt{\lambda_{{\ell}}}-\sqrt{\hat{\lambda}_{{\ell}}} \right)^2- P\right)\nonumber\\&&\hspace{1cm}+\sum_{\ell=1}^L\xi_{\ell}(\gamma_{\ell}-\lambda_{\ell})-\sum_{\ell=1}^L\eta_{\ell}\hat{\lambda}_{\ell}.\label{Lagrange-dual-function-W2}
\end{IEEEeqnarray}
Suppose that the  strict inequality in~\eqref{lambda-hat-positive-W2} does not hold, i.e., $\hat{\lambda}_{\ell}^*=0$. We show that such $\hat{\lambda}_\ell^*$ cannot be the optimal solution to the inner minimization problem.

The Lagrangian term in \eqref{Lagrange-dual-function-W2} depends on $\gamma_{\ell}$ and $\hat{\lambda}_{\ell}$ through the following function: 
\begin{IEEEeqnarray}{rCl}
	G'_{\ell}(\gamma_{\ell},\hat{\lambda}_{\ell})&=&\frac{1}{2}\log\frac{\lambda_{\ell}}{\gamma_{\ell}}+\nu_1\left(\lambda_{\ell}-2\sqrt{\hat{\lambda}_{\ell}(\lambda_{\ell}-\gamma_{\ell})}+\hat{\lambda}_{\ell}\right)\nonumber\\&&\hspace{0.1cm}+\nu_2\left(\sqrt{\lambda_{\ell}}-\sqrt{\hat{\lambda}_{\ell}}\right)^2+\xi_{\ell}(\gamma_{\ell}-\lambda_{\ell})-\eta_{\ell}\hat{\lambda}_{\ell}.\nonumber\\
\end{IEEEeqnarray}
We fix $\gamma_\ell = \gamma_\ell^*$ and then deviate from $\hat{\lambda}_{\ell}^*=0$ to $\hat{\lambda}_{\ell}'=\epsilon$ for some small $\epsilon>0$. The first order change in $G'_\ell(\gamma^*_{\ell},\hat{\lambda}_{\ell})$ can be seen as follows:
\begin{IEEEeqnarray}{rCl}
	&&\hspace{-1cm}G'_{\ell}(\gamma^*_{\ell},\hat{\lambda}^*_{\ell})-G'_{\ell}(\gamma^*_{\ell},\hat{\lambda}'_{\ell})\nonumber\\&=& \nu_1(2\sqrt{\epsilon(\lambda_{\ell}-\gamma_{\ell}^*)}-\epsilon) +\nu_2 (2\sqrt{\lambda_{\ell}\epsilon}-\epsilon)+\eta_{\ell}\epsilon \quad \\
	&=& 2(\nu_2\sqrt{\lambda_{\ell}}+\nu_1\sqrt{\lambda-\gamma^*_{\ell}})\sqrt{\epsilon}+O(\epsilon).\label{Lagrange-dual-final-step-W2}
\end{IEEEeqnarray}
Thus, if $\nu_2 > 0$,  for sufficiently small $\epsilon > 0$, we can strictly decrease $G'_{\ell}(\gamma^*_{\ell},\hat{\lambda}^*_{\ell})$, while satisfying the implicit constraints. This contradicts with the assumption that $\hat{\lambda}^*_{\ell}=0$ is the optimal solution to the inner minimization problem. This proves~\eqref{lambda-hat-positive-W2}. 

Given the strict inequality in~\eqref{lambda-hat-positive-W2}, similar to the KL-divergence metric, we can show that
\begin{IEEEeqnarray}{rCl}
	\gamma_{\ell}^* < \lambda_\ell. \label{gamma-positive-W2}
	\end{IEEEeqnarray}

The strict inequalities in~\eqref{gamma-positive-W2} and~\eqref{lambda-hat-positive-W2} imply that each component has a positive rate, and further $\xi_{\ell}=\eta_{\ell}=0$.  Thus, we can write down the following KKT conditions
\begin{subequations}\label{kkt-conditions-W2}
	\begin{IEEEeqnarray}{rCl}
		\frac{1}{2\gamma^*_{{\ell}}}-\nu_1\sqrt{\frac{{\hat{\lambda}_{{\ell}}^*}}{\lambda_{{\ell}}-\gamma^*_{{\ell}}}}&=&0\quad\label{cond-a-W2}\\
		\nu_1\left(-\sqrt{\frac{\lambda_{{\ell}}-\gamma^*_{{\ell}}}{\hat{\lambda}^*_{{\ell}}}}+1\right)+\nu_2\left(1-\sqrt{\frac{\lambda_{\ell}}{\hat{\lambda}^*_{\ell}}}\right)&=&0\quad\label{cond-b-W2}\\
 \sum_{\ell=1}^L 
		( \lambda_{{\ell}}-2\sqrt{\hat{\lambda}_{{\ell}}^*(\lambda_{{\ell}}-\gamma_{{\ell}}^*)}+\hat{\lambda}_{{\ell}}^*)  &=&D\quad\\
\sum_{\ell=1}^L \left(\sqrt{\lambda_{{\ell}}}-\sqrt{\hat{\lambda}^*_{{\ell}}} \right)^2 &=&P.
	\end{IEEEeqnarray}
\end{subequations}
The derivation of the optimal solution can now be shown as follows.  

Define
\begin{IEEEeqnarray}{rCl}
	\theta_{\ell}&=&\sqrt{\frac{\lambda_{\ell}-\gamma^*_{\ell}}{\hat{\lambda}^*_{\ell}}}.\label{W2-theta-l}
	\end{IEEEeqnarray}
Plugging the above definition into~\eqref{cond-b-W2} yields 
\begin{IEEEeqnarray}{rCl}
		\hat{\lambda}^*_{\ell}&=&\frac{\lambda_{\ell}}{\left(1+\frac{(1-\theta_{\ell})\nu_1}{\nu_2}\right)^2}.\label{W2-hat-lambda-l}
	\end{IEEEeqnarray}
Also, from~\eqref{cond-a-W2}, we get
\begin{IEEEeqnarray}{rCl}
	\gamma^*_{\ell}=\frac{\theta_{\ell}}{2\nu_1}.\label{W2-gamma-l}
	\end{IEEEeqnarray}
Plugging~\eqref{W2-hat-lambda-l} and~\eqref{W2-gamma-l} into~\eqref{W2-theta-l}, we get the following equation:
\begin{IEEEeqnarray}{rCl}
\frac{\theta_{\ell}}{1+\frac{(1-\theta_{\ell})\nu_1}{\nu_2}}=\sqrt{1-\frac{\theta_{\ell}}{2\nu_1\lambda_{\ell}}}.
	\label{eq:theta_fixed_point}
	\end{IEEEeqnarray}
Note that the function $\frac{\theta_{\ell}}{1+\frac{(1-\theta_{\ell})\nu_1}{\nu_2}}$ is an increasing function in $\theta_{\ell}$. Also, the function $\sqrt{1-\frac{\theta_{\ell}}{2\nu_1\lambda_{\ell}}}$ in the interval $\theta_{\ell}\in [0,2\nu_1\lambda_{\ell}]$ is a decreasing function in $\theta_{\ell}$. So, the solution to the above equation is unique.

Thus, $\hat{\lambda}^*_{\ell}$ and $\gamma_{\ell}^*$ in~\eqref{W2-hat-lambda-l} and~\eqref{W2-gamma-l} can be obtained from $\theta_{\ell}$, which is determined via \eqref{eq:theta_fixed_point}.
This proves \eqref{optimal-solution-gamma-W2} and \eqref{optimal-solution-lambda-W2}.

\section{Proof of Corollary~\ref{0-PLF-cor}}\label{extreme-points}

If $P=0$, this falls under the first case in Theorem~\ref{thm-optimal-sol} and Theorem~\ref{cor-W2-optimal-sol}. Here, we have
\begin{IEEEeqnarray}{rCl}
	R(D,0)&=&\frac{1}{2}\sum_{\ell=1}^L\log\frac{\lambda_{\ell}}{\gamma^*_{\ell}(D,0)}.
	\label{eq:R_appendix_H}
\end{IEEEeqnarray}
The perception constraint \eqref{thm-first-clause-P-constraint} and~\eqref{W2-perception-constraint-total}  with $P=0$ implies that $\hat{\lambda}^*_{\ell}(D,0)=\lambda_{\ell}$ for every $\ell\in \{1,\ldots,L\}$. Now, using the expression of optimal $\gamma_\ell^*$ in \eqref{gamma-lambdahat} together with $\hat{\lambda}^*_{\ell}=\lambda_{\ell}$, we have 
\begin{IEEEeqnarray}{rCl}
	\gamma_{\ell}^*(D,0)&=&\frac{2\lambda_{\ell}}{1+\sqrt{1+16\nu_1^2\lambda_{\ell}^2}},
	\label{eq:gamma_appendix_H}
\end{IEEEeqnarray}
where $\nu_1$ is chosen to satisfy the distortion constraint
\eqref{thm-first-clause-D-constraint} and~\eqref{thm-first-clause-D-constraint-W2}, i.e.,
\begin{IEEEeqnarray}{rCl}
	D=\sum_{\ell=1}^L\left(2\lambda_{\ell}-2\sqrt{\lambda_{\ell}(\lambda_{\ell}-\gamma_{\ell}^*(D,0))}\right).
	\label{eq:D_appendix_H}
\end{IEEEeqnarray}
Combining the above proves the desired result. 

\section{Asymptotic Analysis for Perceptually Perfect Reconstruction}

\label{low-rate-proof}

We utilize the optimal solution for the perceptually perfect reconstruction case in Corollary~\ref{0-PLF-cor}, i.e., \eqref{eq:R_appendix_H}, \eqref{eq:gamma_appendix_H} and \eqref{eq:D_appendix_H}. 

\subsection{High-Distortion Compression}
\label{low-0-PLF-app}

Let $D= \left( \sum_{\ell=1}^L 2\lambda_{\ell} \right) - \epsilon$ for some small $\epsilon>0$.
Note that 
by (\ref{eq:D_appendix_H}), this means that we are setting $\epsilon$ to be
\begin{equation}
	\epsilon = \sum_{\ell=1}^L 2 \sqrt{\lambda_{\ell}(\lambda_{\ell}-\gamma^*_{\ell}(D,0))}.
	\label{eq:epsilon_high_distortion}
\end{equation}
In this case, $\gamma_{\ell}^*(D,0)$ should be close to $\lambda_{\ell}$,
and the rate is close to zero.  By (\ref{eq:gamma_appendix_H}), this also
means that $\nu_1$ must be close to zero. 
Then, we can approximate $\gamma_{\ell}^*(D,0)$ as follows:
\begin{IEEEeqnarray}{rCl}
	\gamma_{\ell}^*(D,0)&=&\frac{2\lambda_{\ell}}{1+\sqrt{1+16\lambda_{\ell}^2\nu_1^2}}\nonumber\\
	&=&\frac{\lambda_{\ell}}{1+4\nu_1^2\lambda^2_{\ell}+O(\nu_1^4)}\nonumber\\
	&=&\lambda_{\ell}(1-4\nu_1^2\lambda^2_{\ell})+O(\nu_1^4).
	\label{eq:water-level-high-distortion}
\end{IEEEeqnarray}
Plugging the above into~\eqref{eq:epsilon_high_distortion} yields 
\begin{IEEEeqnarray}{rCl}
	\epsilon = 4\nu_1\sum_{\ell=1}^L\lambda^2_{\ell} + O(\nu_1^2). \label{nu-D-high-0-PLF}
\end{IEEEeqnarray}
The rate expression can now be approximated as follows 
\begin{IEEEeqnarray}{rCl}
	R\left(2\sum_{\ell=1}^L\lambda_{\ell}-\epsilon,0\right)&=&\frac{1}{2}\sum_{\ell=1}^L\log\frac{1+\sqrt{1+16\nu_1^2\lambda^2_{\ell}}}{2}\nonumber\\
	&=&\frac{1}{2}\sum_{\ell=1}^L\log (1+4\nu_1^2\lambda_{\ell}^2+O(\nu_1^4))\nonumber\\
	&=& \frac{1}{2}\sum_{\ell=1}^L 4\nu_1^2\lambda_{\ell}^2+O(\nu_1^4). \label{step2-app}
\end{IEEEeqnarray}
Now, using 
\eqref{nu-D-high-0-PLF} and \eqref{step2-app} to eliminate $\nu_1$, we get 
\begin{IEEEeqnarray}{rCl}
	R\left(2\sum_{\ell=1}^L\lambda_{\ell}-\epsilon,0\right)=\frac{\epsilon^2}{8\sum_{\ell=1}^L\lambda_{\ell}^2}+O(\epsilon^3).
\end{IEEEeqnarray}
To derive the expression for the water level, we use \eqref{nu-D-high-0-PLF} in \eqref{eq:water-level-high-distortion} to get
	\begin{IEEEeqnarray}{rCl}
		&&\gamma^*_{\ell}\left(2\sum_{\ell=1}^L\lambda_{\ell}-\epsilon,0\right)=\lambda_{\ell}-\frac{\epsilon^2\lambda_{\ell}^3}{4\left(\sum_{\ell=1}^L\lambda_{\ell}^2\right)^2}+O(\epsilon^3),\nonumber\\&& \hspace{5cm}\ell\in\{1,\ldots,L\}.
	\end{IEEEeqnarray}

\subsection{Low-Distortion Compression}
\label{high-0-PLF-app}

Let $D=\epsilon$ for some small $\epsilon>0$.
Note that as $\epsilon \rightarrow 0$, we must have $\gamma_\ell^* \rightarrow 0$ by \eqref{eq:D_appendix_H}, and consequently $\nu_1 \rightarrow \infty$ by \eqref{eq:gamma_appendix_H}.

In this regime, we can approximate the 
water levels in \eqref{eq:gamma_appendix_H} as follows
\begin{IEEEeqnarray}{rCl}
	\gamma_{\ell}^*(D,0)&=&\frac{2\lambda_{\ell}}{1+\sqrt{1+16\lambda_{\ell}^2\nu_1^2}}\nonumber\\
	&=&\frac{1}{2\nu_1} -\frac{1}{8\nu_1^2\lambda_{\ell}}+
	O\left(\frac{1}{\nu_1^3}\right).\;\;\;\label{eq:sub}
\end{IEEEeqnarray}
Plugging \eqref{eq:sub} into the distortion constraint \eqref{eq:D_appendix_H}, we have
\begin{IEEEeqnarray}{rCl}
		\epsilon&=& \sum_{\ell=1}^L\left(2\lambda_{\ell}-2\sqrt{\lambda_{\ell}\left(\lambda_{\ell}-\gamma_{\ell}^*(D,0)\right)}\right)\nonumber\\ 
		&=& \frac{L}{2\nu_1}-\frac{1}{16\nu_1^2}\sum_{\ell=1}^L\frac{1}{\lambda_{\ell}}+O\left(\frac{1}{\nu_1^3}\right),
\end{IEEEeqnarray}
which implies
	\begin{IEEEeqnarray}{rCl}
		\frac{1}{\nu_1}=\frac{2\epsilon}{L}+\frac{\epsilon^2}{2L^3}\sum\limits_{\ell=1}^L\frac{1}{\lambda_{\ell}}.\label{approximation1}	
\end{IEEEeqnarray}
Substituting \eqref{approximation1} into \eqref{eq:sub} shows that
the water levels in the low-distortion regime are given by
\begin{IEEEeqnarray}{rCl}
		\gamma_{\ell}^*(\epsilon,0)
		&=& \frac{\epsilon}{L}-\frac{\epsilon^2}{2L^2\lambda_{\ell}}+\frac{\epsilon^2}{4L^3}\sum_{\ell=1}^L\frac{1}{\lambda_{\ell}}+O(\epsilon^3),\nonumber\\&&\hspace{3cm} \ell\in\{1,\ldots,L\}.\label{low-D-P0-water-level-approx}
\end{IEEEeqnarray}
The rate expression can now be approximated as follows
\begin{IEEEeqnarray}{rCl}
		&&\hspace{-0.5cm}R(\epsilon,0)\nonumber\\&=& \frac{1}{2}\sum_{\ell=1}^L\log \frac{\lambda_{\ell}}{\gamma^*_{\ell}(\epsilon,0)}\\
		&=&\frac{1}{2}\sum_{\ell=1}^L\log \frac{L\lambda_{\ell}}{\epsilon}\nonumber\\&&\hspace{0cm}-\frac{1}{2}\sum\limits_{\ell=1}^L\log\Bigg(1
		-\frac{\epsilon}{2L\lambda_{\ell}}+ \frac{\epsilon}{4L^2}\sum_{\ell'=1}^L\frac{1}{\lambda_{\ell'}}
		+O(\epsilon^2)\Bigg)\quad\;\;\;\\
		&=&\frac{1}{2}\sum_{\ell=1}^L\log \frac{L\lambda_{\ell}}{\epsilon}
		-\frac{1}{2}\sum\limits_{\ell=1}^L\left(
		-\frac{\epsilon}{2L\lambda_{\ell}}
		+\frac{\epsilon}{4L^2}\sum_{\ell'=1}^L\frac{1}{\lambda_{\ell'}}
		\right)
		\nonumber\\&&\hspace{5.5cm}+O(\epsilon^2)\\
		&=& \frac{1}{2}\sum_{\ell=1}^L\log \frac{L\lambda_{\ell}}{\epsilon}+
		\frac{\epsilon}{8L}\sum_{\ell=1}^L\frac{1}{\lambda_{\ell}}+O(\epsilon^2).
\end{IEEEeqnarray}
This concludes the proof.

\bibliographystyle{IEEEtran}
\bibliography{JSAIT_references}

\begin{thebibliography}{10}
\providecommand{\url}[1]{#1}
\csname url@samestyle\endcsname
\providecommand{\newblock}{\relax}
\providecommand{\bibinfo}[2]{#2}
\providecommand{\BIBentrySTDinterwordspacing}{\spaceskip=0pt\relax}
\providecommand{\BIBentryALTinterwordstretchfactor}{4}
\providecommand{\BIBentryALTinterwordspacing}{\spaceskip=\fontdimen2\font plus
\BIBentryALTinterwordstretchfactor\fontdimen3\font minus
  \fontdimen4\font\relax}
\providecommand{\BIBforeignlanguage}[2]{{%
\expandafter\ifx\csname l@#1\endcsname\relax
\typeout{** WARNING: IEEEtran.bst: No hyphenation pattern has been}%
\typeout{** loaded for the language `#1'. Using the pattern for}%
\typeout{** the default language instead.}%
\else
\language=\csname l@#1\endcsname
\fi
#2}}
\providecommand{\BIBdecl}{\relax}
\BIBdecl

\bibitem{image-comp1}
E.~Agustsson, M.~Tschannen, F.~Mentzer, R.~Timofte, and L.~Van~Gool,
  ``Generative adversarial networks for extreme learned image compression,'' in
  \emph{Proc. IEEE Conf. Comput. Vis. Pattern Recognit. (CVPR)}, 2019, pp.
  221--231.

\bibitem{image-comp2}
J.~Ball\'e, V.~Laparra, and E.~P. Simoncelli, ``End-to-end optimized image
  compression,'' in \emph{Proc. Int. Conf. Learn. Represent. (ICLR)}, 2017.

\bibitem{image-comp3}
L.~Theis, W.~Shi, A.~Cunningham, and F.~Husz\'ar, ``Lossy image compression
  with compressive autoencoders,'' in \emph{Proc. Int. Conf. Learn. Represent.
  (ICLR)}, 2017.

\bibitem{image-comp4}
F.~Mentzer, E.~Agustsson, M.~Tschannen, R.~Timofte, and L.~V. Gool,
  ``Conditional probability models for deep image compression,'' in \emph{Proc.
  IEEE Conf. Comput. Vis. Pattern Recognit. (CVPR)}, 2018.

\bibitem{blau2019rethinking}
Y.~Blau and T.~Michaeli, ``Rethinking lossy compression: The
  rate-distortion-perception tradeoff,'' in \emph{Proc. ACM Int. Conf. Mach.
  Learn. (ICML)}, 2019, pp. 675--685.

\bibitem{Saldi}
N.~Saldi, T.~Linder, and S.~Y\"{u}ksel, ``Output constrained lossy source
  coding with limited common randomness,'' \emph{IEEE Trans. Inf. Theory},
  vol.~61, no.~9, pp. 4984--4998, 2015.

\bibitem{Theis-Wagner}
L.~Theis and A.~Wagner, ``A coding theorem for the rate-distortion-perception
  function,'' in \emph{Int. Conf. Learn. Represent. (ICLR), Neural Compress.
  Workshop}, 2021.

\bibitem{LiElGamal}
C.~T. Li and A.~El~Gamal, ``Strong functional representation lemma and
  applications to coding theorems,'' \emph{IEEE Trans. Inf. Theory}, vol.~64,
  no.~11, pp. 6967--6978, 2018.

\bibitem{Jun-Ashish2021}
G.~Zhang, J.~Qian, J.~Chen, and A.~Khisti, ``Universal
  rate-distortion-perception representations for lossy compression,'' in
  \emph{Proc. Adv. Neural Inf. Process. Syst. (NeurIPS)}, 2021, pp.
  11\,517--11\,529.

\bibitem{wagner2022rate}
A.~B. Wagner, ``The rate-distortion-perception tradeoff: The role of common
  randomness,'' \emph{arXiv:2202.04147}, 2022.

\bibitem{Jun-JSAIT}
J.~Chen, L.~Yu, J.~Wang, W.~Shi, Y.~Ge, and W.~Tong, ``On the
  rate-distortion-perception function,'' \emph{IEEE J. Sel. Areas Inf. Theory},
  vol.~3, no.~4, pp. 664--673, 2022.

\bibitem{freirich2021theory}
D.~Freirich, T.~Michaeli, and R.~Meir, ``A theory of the distortion-perception
  tradeoff in wasserstein space,'' \emph{Proc. Adv. Neural Inf. Process. Syst.
  (NeurIPS)}, vol.~34, pp. 25\,661--25\,672, 2021.

\bibitem{yan2021perceptual}
Z.~Yan, F.~Wen, R.~Ying, C.~Ma, and P.~Liu, ``On perceptual lossy compression:
  The cost of perceptual reconstruction and an optimal training framework,'' in
  \emph{Proc. ACM Int. Conf. Mach. Learn. (ICML)}, 2021, pp. 11\,682--11\,692.

\bibitem{Huan-Liu}
H.~Liu, G.~Zhang, J.~Chen, and A.~Khisti, ``Lossy compression with distribution
  shift as entropy constrained optimal transport,'' in \emph{Proc. Int. Conf.
  Learn. Represent. (ICLR)}, 2022.

\bibitem{Jun-Ashish2023}
S.~Salehkalaibar, B.~Phan, J.~Chen, W.~Yu, and A.~Khisti, ``On the choice of
  perception loss function for learned video compression,'' in \emph{Proc. Adv.
  Neural Inf. Process. Syst. (NeurIPS)}, 2023.

\bibitem{Cover}
T.~M. Cover and J.~A. Thomas, \emph{Elements of Info. Theory, 2nd Ed}.\hskip
  1em plus 0.5em minus 0.4em\relax Wiley, 2006.

\bibitem{Jun-differential-entropy}
L.~Song, J.~Chen, and C.~Tian, ``Broadcasting correlated vector gaussians,''
  \emph{IEEE Trans. Inf. Theory}, vol.~61, no.~5, pp. 2465--2477, 2015.

\end{thebibliography}

\end{document}